\documentclass[prb,aps,floatfix,amsmath,amssymb,superscriptaddress]{revtex4}

\usepackage{mathtools}
\usepackage{amsfonts}
\usepackage{amssymb}

\usepackage{graphicx}
\usepackage[all]{xy}

\usepackage{amsthm}

\newtheorem{theorem}{Theorem}[section]
\newtheorem{lemma}[theorem]{Lemma}
\newtheorem{definition}[theorem]{Definition}

\newcommand{\mO}{{\mathcal O}}

\newcommand{\Ls}{L_{\rm sys}}

\newcommand{\ZZ}{\mathbb{Z}}

\newcommand{\be}{\begin{equation}}
\newcommand{\ee}{\end{equation}}

\begin{document}
\title{Classifying Quantum Phases With The Torus Trick}
\author{M. B. Hastings}
\affiliation{Microsoft Research, Station Q, CNSI Building, University of California, Santa Barbara, CA, 93106}
\affiliation{Quantum Architectures and Computation Group, \\
Microsoft Research, Redmond, WA 98052 USA}

\begin{abstract}
Classifying phases of local quantum systems is a general problem that includes special cases such as free fermions, commuting projectors, and others.
An important distinction in this classification should be made between classifying periodic and aperiodic systems.  A related distinction is that between homotopy invariants (invariants which remain constant so long as certain general properties such as locality, gap, and others hold) and locally computable invariants (properties of the system that cannot change from one region to another without producing a gapless edge between them).  We attack this problem using a technique
inspired by Kirby's ``torus trick" in topology.
We use this trick to reproduce results for free fermions (in particular, using the trick to reduce the aperiodic classification to the simpler problem of periodic classification).  We also show that a similar trick works for interacting phases which are nontrivial but lack anyons; these results include symmetry protected phases.  A key part of this work is an attempt to classify quantum cellular automata (QCA).
\end{abstract}
\maketitle
The problem of classifying different phases of quantum Hamiltonians is a major problem in condensed matter physics and quantum information today.  There are many different forms of this problem, depending upon the specific kind of system being classified.  The general form of the problem starts by defining some property $*$
of Hamiltonians, where property $*$ typically includes properties such as a spectral gap and spatially local interactions on some finite-dimensional lattice,
and potentially also includes various symmetries including group symmetries or time reversal symmetry.  We say that two Hamiltonians $H_0$ and $H_1$ with property $*$ are in the same phase if we can find a continuous path of Hamiltonians $H_s$, with $0 \leq s \leq 1$, connecting $H_0$ to $H_1$, with all Hamiltonians in this path having property $**$, where $**$ is some property related to $*$, possibly with slightly relaxed locality properties as discussed below.

Various results along this line have been obtained with particular success in the case that $*$ refers to gapped, local, noninteracting fermion Hamiltonians with various symmetry properties.  A full classification is now known\cite{kitaev,ludwig} in this case, in all dimensions and all symmetry classes.  More recently, there has been much interest in studying symmetry protected phases of interacting systems\cite{spt,sptU}; most of that work is restricted to the case in which anyons are not present and in this paper in fact the absence of anyons will play a very important role in our techniques as defined more precisely later.

Often we are interested in classification only up to {\it stable equivalence}, as emphasized by Ref.~\onlinecite{kitaev}.  In this case, we define certain systems to be {\it trivial}.  These will be systems in which the Hamiltonian $H_{triv}$ is a sum of terms on different sites, with no coupling between sites.
We also need to define a method of adding two systems together.  In the non-interacting case, the Hamiltonian is simply a matrix with basis elements corresponding to sites, and the sum of two systems is simply the direct sum of the two matrices, while in the interacting case, to add together two systems with Hamiltonians $H$ and $H'$, we take the tensor product of their Hilbert spaces, and the Hamiltonian of the combined system is $H \otimes I + I \otimes H'$.
Then, we say that $H$ and $H'$ are in the same phase if we can find two trivial Hamiltonian $H_{triv}$ and $H'_{triv}$ such that
$H \otimes I + I \otimes H_{triv}$ and $H' \otimes I +I \otimes H'_{triv}$ are connected by a continuous path of Hamiltonians with the given property $*$ or $**$.

There are several distinctions that are important in this classification.  One is the distinction between classifying finite or infinite systems, while another is the distinction between periodic and aperiodic systems.  A final distinction is that between homotopy invariants and locally computable invariants.  In the next three subsections, we explain these distinctions, using the case of the free fermion classification problem to exemplify them.

We will use the torus trick in an attempt to remove the distinction between periodic and aperiodic systems, using it to show for certain types of quantum systems that given an aperiodic quantum system with some given property (such as a gap, local interactions, symmetry etc...) and given some set  of sites, there is a {\it periodic} system that agrees with the original system on that set and has the same or similar property.
  In some cases, such as free fermions, we are able to reduce the aperiodic classification to the periodic classification and then use results on the periodic classification (in this case, the K-theory of vector bundles) to classify aperiodic systems.  Given that the trick is useful in this one area, of course one is motivated to look for other areas to apply it.
 For other types of Hamiltonians with anyonic degrees of freedom we find troubles with a straightforward application of the torus trick in section \ref{complications}.  However, we find that for a type of system called a quantum cellular automaton\cite{QCA} (QCA) the torus trick can be applied in any dimension.  We will motivate these QCA by presenting a possible formal definition of systems which are nontrivial but have no intrinsic topological order in terms of QCA in section \ref{nointo}; using this definition the torus trick can be applied to such systems without intrinsic topological order.  Finally, we present some partial results on a classification of QCA in higher dimensions with symmetry (the case of no symmetry in one dimensions is in Ref.~\onlinecite{QCA}).
Since our main focus in this paper is developing the torus trick for quantum systems, much of our results on classifying QCA and systems without intrinsic topological order are only partial and fuller results will be given elsewhere.

Some comments on notation: we consider lattice systems with sites labeled by some index $i$, with each site corresponding to a point in some space $M$ called the ambient space.
We use ${\rm diam}(...)$ to denote the diameter of a set and ${\rm dist}(i,j)$ to denote the distance between two sites $i$ and $j$ and ${\rm dist}(X,Y)$ to denote the distance between two sets $X,Y$ defined as ${\rm min}_{i\in X,j \in Y}({\rm dist}(i,j))$.
We use $\Vert ... \Vert$ to denote the operator norm.
The ambient space may be finite or infinite.
While general choices of $M$ are possible, in this paper 
we consider only the cases of $R^d$ or $T^d$ and we will use a Euclidean metric throughout.

\subsection{Finite vs. Infinite Lattice Systems}
We take $M=R^d$ for an infinite system, or $M=T^d$, the $d$-dimensional torus, to get a finite system.  We will bound the number of sites so that the number of sites within distance $r$ of any point is bounded by
\be
O(r+1)^d,
\ee
where $O(...)$ is big-O notation.
For a finite system with $M=T^d$, we choose to parametrize the torus by $d$ numbers $x_a$, with $0 \leq x_a < \Ls$, where $\Ls$ is the ``system size", and we use the usual Euclidean metric with this parametrization.  This will seem very natural to physicists but may be a less natural parametrization for others; the reason for this is that we will fix the length scale for the distance between sites to be of order $1$, and we will have the interactions decay on some length scale $R$ (which may be much larger than $1$); then we will be interested in the case $\Ls>>R$, and we will obtain bounds that are uniform in $\Ls$.

For free fermions,
the Hilbert space has a finite dimensional Hilbert space on each lattice site (we allow more than one state per site) and the whole Hilbert space is the direct sum of these Hilbert spaces, while for interacting systems, there is some finite dimensional space on each site and the whole Hilbert space is the tensor product of these sites.
For free fermions,
the Hamiltonian $H$ is a Hermitian matrix with some locality property on the matrix elements.  We regard $H$ as a block matrix, with one block per site.  One possibility is to require that the block $H_{ij}$ coupling site $i$ to site $j$ be exponentially small in the distance between sites $i$ and $j$, with some length scale $R$ setting the decay rate; another possibility is to require that the matrix elements are strictly zero beyond some finite range $R$.  We refer to this length scale $R$ as the {\it range} of the interactions.
We require that $H$ have a gap in its spectrum.  For simplicity, let us fix this gap near $0$, requiring that the spectrum not contains any energies smaller in absolute value than $\Delta E$, for some given $\Delta E$.   Finally, we bound the norm of the terms in $H$ in some way later.
In addition to these requirements, there may be some symmetry properties imposed on $H$, such as time-reversal symmetry (either with or without spin-orbit coupling) and so on.

In the case of an infinite system of free fermions, an interesting classification problem is: given $H_0$ and $H_1$ with property $*$, we ask for a continuous path of Hamiltonians $H_s$ with property $*$ for $0 \leq s \leq 1$, where property $*$ is the property that $R$ is finite (either in the case of exponential decay or strictly finite range for the matrix elements), that $\Delta E$ is positive, and that $H$ has any desired symmetry properties.

However, if the lattice has a finite number of sites, then the question as phrased above is uninteresting.  For one thing, we simply required that $R$ is finite in the previous paragraph.  However, if the system has some finite size $\Ls$, then given any two Hamiltonians, $H_0$ and $H_1$, each with spectral gap $\Delta E$, there always is a path $H_s$ connecting them which preserves a gap of at least $\Delta E$ and which obeys the required symmetries for sufficiently large $R$ (i.e., pick $R$ large enough compared to $\Ls$).  So, for a finite system it is necessary to consider analytic details as to the magnitude of $R$.  Similarly it is also necessary to consider the magnitude of $\Delta E$, rather than simply requiring that $\Delta E$ be non-zero.  So for a finite system, the relevant classification problem is: given $H_0$ and $H_1$ with property $*$, we ask for a continuous path of Hamiltonians $H_s$ with property $**$ for $0 \leq s \leq 1$, where $*$ is a given lower bound $\Delta E$ on the gap and a given upper bound $R$ on the range of the interactions as well as any symmetry requirements and where $**$ has some other lower bound $\Delta E'$ on the gap and other upper bound $R'$ on the range of the interactions as well as any symmetry requirements.

Results showing the existence of such a path $H_s$ will be most interesting if they give $\Delta E'$ and $R'$ that are independent of the size of the lattice and depend only upon the original $\Delta E$ and $R$.  In this paper, we will largely focus on such results for finite systems, both for free fermions and for interacting systems; that is, we derive quantitative bounds (though we do not worry about constant factors).  Later, we discuss the distinction between infinite and finite systems for QCA; in this case, the infinite case requires some care to define.

\subsection{Periodic vs. Aperiodic Systems and The Torus Trick}
Another important distinction is that between periodic and aperiodic systems.  We first explain this distinction in the case of free fermions.  We label the sites by coordinates $n_1,...,n_d$.  We choose $d$ linearly independent $d$-dimensional vectors $v_a$, for $a=1,...,d$, with the set of sites invariant under translation by these vectors.
Then, we say that $H$ is periodic if
$H_{ij}=H_{kl}$ if $i+v_a=k$ and $j+v_a=l$, where $i+v_a$ denotes the site obtained by translating site $i$ by vector $v_a$.

Note that these vectors $v_a$ need not be basis vectors for the lattice.  For example, if we work on a $2$-dimensional square lattice with sites at integer coordinates, we might choose that $v_1=(L,0)$ and $v_2=(0,L)$ for some integer $L>1$.  Note also that if a system is periodic for a given choice of $v_a$, then it is periodic for any choice $n_a v_a$ for any integers $n_a$.  Informally, we can ``increase the size of the unit cell".

For other systems, such as Hamiltonians for interacting quantum systems or for QCA, the notion of a periodic system still makes sense as a Hamiltonian or QCA which commutes with translation operators.    For a finite system, the translation operators are unitaries $T_a$ for $a=1,...,d$ which  translate the system on the lattice by the vector $v_a$, while for an infinite system they are defined as algebra automorphisms (see the definition of QCA later).

So, two distinct problems are the classification of periodic or aperiodic systems.  As explained below, periodic systems can be classified using results from K-theory on the classification of vector bundles.  However, the classification of aperiodic systems is much more difficult.  One approach to this is using controlled K-theory\cite{cKt}.  Another approach is based on a result of Kitaev's that maps an aperiodic lattice free fermion system to a Dirac Hamiltonian with a smoothly varying mass term, called a texture\cite{kitaev}.

In this paper we adopt a third approach to solving this problem. This approach may have some advantages: it gives more quantitative bounds than the controlled K-theory results with which I am familiar (however, I am fairly unfamiliar with that literature, so it is possible that similar bounds are available there).  It may also be  simpler than the method of textures of Kitaev, which relies on a theorem stated in Ref.~\onlinecite{kitaev} without proof.  However, the main reason we introduce this technique is that it will also have
application to certain interacting systems.

Our approach is inspired by the {\it torus trick}, a technique invented by Kirby\cite{kirby} in 1968 to solve several problems in topology.  The techniques in the present paper are self-contained, so that no previous knowledge of this trick is required to read it.  The general kind of results here will roughly have the following form: given some aperiodic Hamiltonian $H$ with some property $*$ and given some set $Z\subset M$, there is a {\it periodic} Hamiltonian $H'$ with property $**$ such that $H$ and $H'$ agree on $Z$.
This style of result is very similar to the original application of the torus trick, where instead of considering Hamiltonians or QCA, the trick was applied to homeomorphisms from $R^d$ to $R^d$ (in this context, the analogue of a periodic homeomorphism is a homeomorphism $f$ such that $f(x+v_a)=v_a+f(x)$ for some set of vectors $v_a$).  The property $**$ will often be very similar to $*$ except for some slight weakening of the locality properties.  Ideally, if $Z$ has diameter $L$, then the periodicity of $H'$ will be on a scale only slightly larger than $L$ (that is, the basis vectors $v_a$ used to define translation should have the property that $|v_a|$ is comparable to $L$).

We emphasize here that $Z$ is a subset of $M$ rather than a set of sites; then, $H$ and $H'$ agree on $Z$ if they agree on the sites in $Z$.  The reason to specify that $Z$ is a subset of $M$ is that this will be useful later for certain stronger results.  In some cases, we will be able to show, for example, that if $Z$ is a hypercube with each side having length $L$ and with the center of $Z$ being at coordinate $z$, then $H'$ depends smoothly on $H$ and $z$ (more precisely, it will depend smoothly away from certain discontinuities; however, we will show a stable equivalence of $H'$ at points near the discontinuity).  By choosing $Z$ to be a subset of $M$, this makes it easier to talk about changing $z$ continuously.  We often write $H'(H,Z)$ to emphasize that $H'$ is a function of $H$ and $Z$.

Having derived results of this form, we will then apply them in combination with results on the classification of periodic systems to the specific classification of the aperiodic system at hand.  The particular application of this will depend on the quantum system we consider.

\subsection{Homotopy Invariants vs. Local Invariants}
An interesting final relation is that between homotopy invariants and locally computable invariants.  This distinction was highlighted in Ref.~\onlinecite{QCA} for a classification of one-dimensional QCA, but we discuss it here in generality.
By ``homotopy invariant", we mean the kind of classification problem discussed above: is there a continuous path $H_s$ connecting $H_0$ to $H_1$, possibly with stabilization?

``Local invariant" refers to a different but related question.  We consider two Hamiltonians, $H_0$ and $H_1$, with property $*$ and pick two sets $Z_0$ and $Z_1$ with the distance between $Z_0$ and $Z_1$ being large.  Then, we ask if there is some Hamiltonian $H$, with some property $**$ such that $H$
agrees with $H_0$ on $Z_0$ and $H$agrees with $H_1$ on $Z_1$.  This is a question of whether there is a system $H$ that interpolates between two different systems $H_0$ and $H_1$.  A local invariant is some quantity that could be obtain from $H_0$ on $Z_0$ or $H_1$ on $Z_1$ that is an obstruction to finding such an interpolation.

The torus trick, especially the ``stronger result" mentioned in the previous section (that $H'(H,Z)$ could depend smoothly upon $H$ and $z$ up to some stable equivalence at discontinuities) will be useful in relating homotopy invariants to local invariants as follows.  If there is a Hamiltonian $H$ that interpolates as desired, and if such a stronger result held, then we obtain
a continuous path of periodic Hamiltonians from $H'(H_0,Z_0)$ to $H'(H_1,Z_1)$ (again up to stable equivalence).  So, if one can interpolate between $H_0$ on $Z_0$ and $H_1$ on $Z_1$, then $H'(H_0,Z_0)$ and $H'(H_1,Z_1)$ are homotopy equivalent.  In some cases, the converse will also be true (homotopy equivalence of $H'(H_0,Z_0)$ and $H'(H_1,Z_1)$ will imply that we can find an interpolating Hamiltonian).
For free fermion systems, this relation holds.
However, this is not generally true: we will explain a case later of an interacting system where one can interpolate between two periodic Hamiltonians which are not homotopy equivalent.  This will be an example of a system with anyons, and the existence of this case is one reason that in this paper we focus on systems which lack anyons.

\section{The Torus Trick for Free Fermions}
After these generalities, we now explain the torus trick in the case of free fermions, before later studying interacting systems.
Given an aperiodic Hamiltonian $H$, we construct a periodic Hamiltonian $H'(H,Z)$ which agrees with $H$ on some set $Z$.  Then, we apply this result to classify aperiodic systems.

In this section, we explain only the case of free fermion Hamiltonians with no superconductivity or time-reversal symmetry or sublattice symmetry.  This is class A in the 10-fold way classification\cite{10fold}.  However, it is easy to see that for any of the other $9$ classes, the symmetry ``goes along for the ride"; that is, while we start with an aperiodic Hamiltonian in class A and construct a periodic Hamiltonian in class A, the same construction starts with an aperiodic Hamiltonian in any given class and constructs a periodic Hamiltonian in the same class.  There are some details that the reader can verify in this, namely that one can preserve the symmetry while ``healing the puncture"(defined later) and also that one can ``join" (defined later) two Hamiltonians while preserving the symmetry.  In classes with sublattice symmetry, the subspace on each site should be defined to be the direct sum of two subspaces of the same dimension, corresponding to the two different sublattices, rather than having different sublattices located at different points in space, and in classes with time-reversal or particle-hole symmetry, the corresponding symmetry operation should be block-diagonal so that, for example, time-reversed pairs for a spin-$1/2$ system are both on the same site rather than on different sites.  This is done so that there exists a gapped block-diagonal Hamiltonian, with blocks corresponding to sites, which respects the symmetry.  So, all our results apply to all $10$ classes.

The trick is applicable to finite or infinite systems.  We explain it with ambient space $M=R^d$ but to apply it to finite spaces with ambient space $M=T^d$ with linear size $\Ls$, we replace the immersion of the punctured torus in $R^d$ below by an immersion in a subset of $T^d$ of diameter sufficiently smaller than $\Ls$.

We impose an exponential decay on the terms in $H_{ij}$ by requiring that, for all sites $i$,
\be
\label{decay}
\sum_{j, {\rm dist}(i,j)\geq r} \Vert H_{ij} \Vert \leq J \exp(-r/R),
\ee
for some positive constants $J,R$.  This bound implies a bound on the norm $\Vert H \Vert$ by $J$ as it bounds $\sum_j \Vert H_{ij}\Vert \leq J$ and this row bound on $H$ bounds the norm of $H$.  Recall that there may be more than one state per site so that $H_{ij}$ may be a matrix rather than a scalar.  The particular form of the bound is not too important; for example, if we set to zero all terms $H_{ij}$ with ${\rm dist}(i,j)>r$ then this produces only an exponentially small change in $H$ and so for sufficiently large $r$ it does not close the gap.

\subsection{Constructing a Periodic Hamiltonian With The Torus Trick}
The construction in this section proves the following:
\begin{theorem}
\label{fftorus}
Consider a free fermion Hamiltonian obeying Eq.~(\ref{decay}) with gap $\Delta E$.
Then, for any hypercube $Z$ of linear size $L$, for $L$ sufficiently large compared to $RJ/\Delta E$, there is a periodic Hamiltonian
$H'(H,Z)$ that agrees with $H$ on $Z$ with $H'$ obeying Eq.~(\ref{decay}) for some different $R'$ and $H'(H,Z)$ having gap $\Delta E'$ with $R'$ upper bounded by a constant times $R$ and $\Delta E'$ lower bounded by a positive constant times $\Delta E$.
These constants depend on dimension.

The periodicity of the Hamiltonian is defined by translation in the directions of the axes of the hypercube by distance $2\pi L$.

There also exists a Hamiltonian $H'_{torus}(H,Z)$ defined on a torus of linear size $2\pi L$ which obeys Eq.~(\ref{decay}), such that if we unfurl (as described below) $H'_{torus}(H,Z)$ we obtain $H'(H,Z)$ and such that $H'_{torus}(H,Z)$ also obeys similar bounds on its decay rate and gap in terms of $R$ and $\Delta E$ up to constant factors.
\end{theorem}
While we stated that translation is by distance $2\pi L$ and the size of the torus is $L$, one can choose those distances to be any value larger than $L$ (the value chosen will determine some of the constants in the above theorem).  This is useful if the ambient space is a torus as only certain periodicities can be fit within the torus.

\begin{figure}
\includegraphics[width=3in]{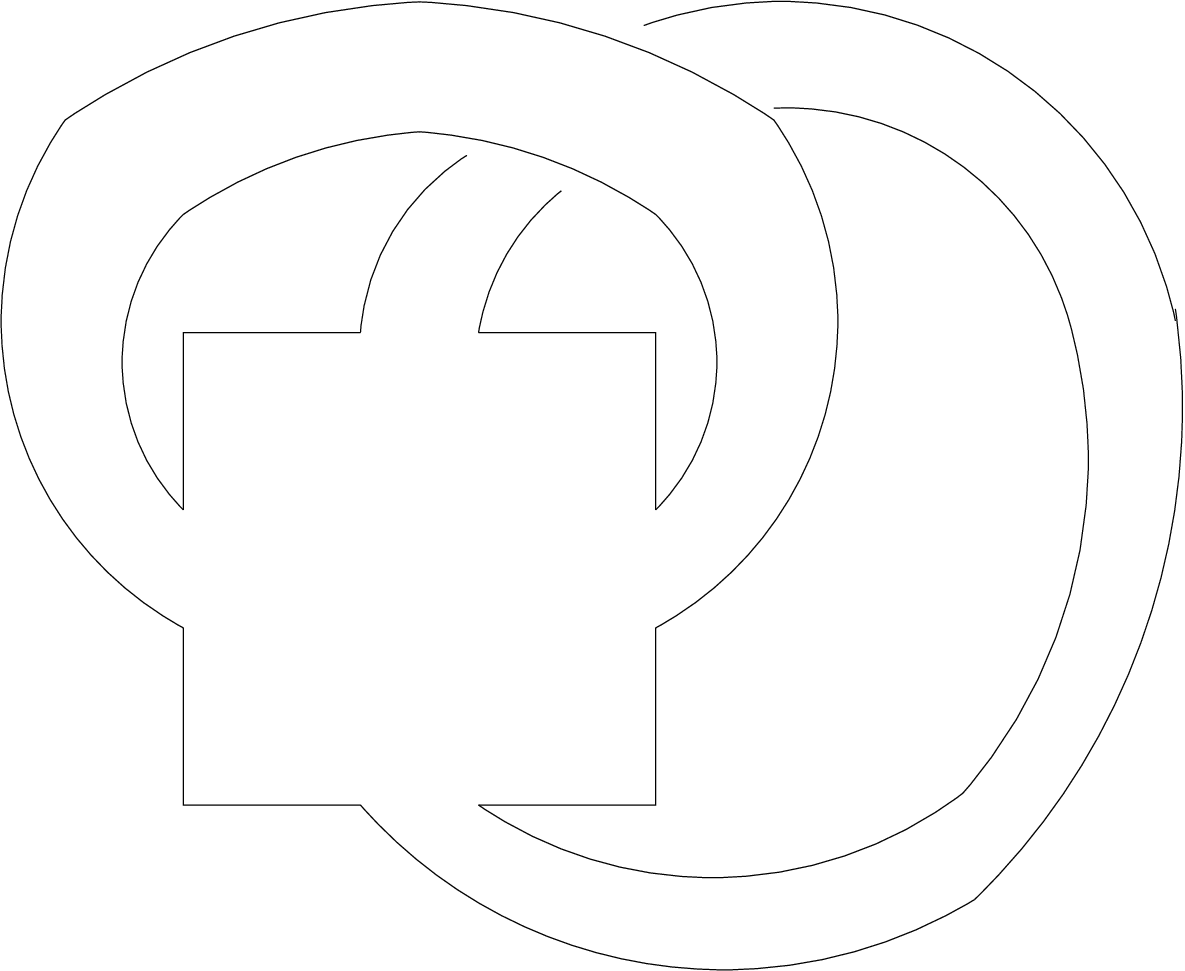}
\caption{Immersed punctured torus in two dimensions}
\label{FigImmerse}
\end{figure}

The key idea in the torus trick is that one can immerse a punctured $d$-dimensional torus in $R^d$.  An example immersion is shown in two dimensions in Fig.~\ref{FigImmerse} (physicists may recognize the figure as being very similar to a Hall bar with source and drain joined).  In general, in $d$ dimensions the immersion is constructed by embedding $d$-different copies of $T^{d-1} \times [0,1]$, with certain restrictions on the intersections.

The torus trick involves the following steps: first, pullback the Hamiltonian in $R^d$ to a Hamiltonian on the punctured torus.  We explain this pullback below; taking $L$ sufficiently large compared to $R$ is important to define the pullback.  This Hamiltonian on the punctured torus will have roughly the same locality as the original Hamiltonian (the interaction range may be slightly increased) but it may not have a gap.  So, the next step is to restore the gap by ``healing the puncture".  A key part of this step will be that, in some sense, the Hamiltonian will still be gapped away from the puncture so that we can heal the puncture just by modifying the Hamiltonian near the puncture so that locality is not violated.  
 Physically, this is familiar from the quantum Hall effect, where a puncture supports gapless edge modes but the bulk remains gapped; see lemma \ref{localfn} and Eq.~(\ref{bulkgap}).
This property that we can heal the puncture just by modifying the system near the puncture determines to which systems we can apply the torus trick: we will see in section \ref{nointo} that such healing is not possible for certain interacting systems with intrinsic topological order, for example.  Then, having healed the puncture, the last step is to ``unfurl" the Hamiltonian to generate a periodic Hamiltonian $H'(H,Z)$.

We begin by defining the immersion.
 Parametrize the punctured torus by angles $\theta_1,...,\theta_d$, all in the range $0 \leq \theta_a < 2\pi$.
Let us use the Euclidean metric to measure distances in both $R^d$ and $T^d$.
Let $f$ be some fixed immersion from the punctured torus to $R^d$.  We will pick $f$ so that the immersion is contained within a hypercube of linear size $2$ centered at the origin.  The line in Fig.~(\ref{FigImmerse}) is mapped back to a single point, the puncture.
Note that the map is not one-to-one.  However we pick $f$ so that the image of the punctured torus contains a hypercube of linear size $1$ centered at the origin, with all points in that hypercube having a unique inverse.  We choose the inverse map on those points to be: a point $z$ with coordinates $x_1,...,x_d$ is mapped to $\theta_1,...,\theta_d$ with $\theta_a=x_a$.  Thus, the points in that hypercube are mapped back without any distortion of angles. All points in this hypercube map under $f^{-1}$ to points with some nonzero distance from the puncture (thus, this hypercube does not extend to the boundary of the ``square" in Fig.~\ref{FigImmerse}).
Finally, we pick $f$ so that its inverse does not ``stretch" distances too much for nearby points; more precisely, we pick $f$ so that for any two points, $x$ and $y$ in the pre-image with ${\rm dist}(x,y) \leq c_{inj}$ for some constant $c_{inj}$ of order unity, we have
\be
\label{stretch}
{\rm dist}(x,y) \leq c_{inj} \; \rightarrow \;
{\rm dist}(f(x),f(y)) \geq C {\rm dist}(x,y),
\ee
where $C>0$ is some constant of order unity, and ${\rm dist}(x,y)$ is the distance using the Euclidean metric.
Note that this means that
$f$ is injective on any set in the pre-image of diameter smaller than $c_{inj}$.  Note also that this bound Eq.~(\ref{stretch}) cannot hold for all $x,y$ as then $f$ would be injective everywhere.

We will define a family of functions $f_{z,L}$ by
\be
f_{z,L}(x)=z+Lf(x).
\ee
 Then, any point in the hypercube of linear size $L$ centered at $z$ has a unique inverse under $f_{z,L}$.
Then, for a given choice of set $Z$ we will define the immersion by such a function $f_{z,L}$, with $z$ and $L$ such that $Z$ is contained in the hypercube of linear size $L$ centered at $z$.
The particularly simple form of $f^{-1}(x)$ in the hypercube near the origin ($\theta_a=x_a$ for such $x$) will ensure that the periodic Hamiltonian indeed agrees with $H_0$ on $Z$.
Eq.~(\ref{stretch}) will play a key role in ensuring locality of interactions in the Hamiltonian pulled back to the punctured torus.

To simplify some of the statements below, we now re-parametrize the punctured torus so that it is parametrized by $d$ coordinates ranging from $0$ to $2\pi L$; having done this, the function $f^{-1}_{z,L}(x)$ does not distort distances for points in $Z$.  Another advantage of this parametrization is that we still have $O((r+1)^d)$ sites within distance $r$ of any point on the punctured torus.

We now define $H_{pt}$, which is the pullback of Hamiltonian $H$ to the punctured torus.  The set of sites on the punctured torus will be the pre-image of the set of sites in the image of the immersion; that is, for every point in the image which contains a site, all the points in the pre-image of that point will also contain a site.  Note that since the immersion is not one-to-one, a given site in the image might correspond to several sites in the pre-image.
In an abuse of notation, if a site $\tilde i$ in the pre-image corresponds to some site $i$ in the image we write $f(\tilde i)=i$.
Then, given two sites in the pre-image, called $\tilde i$ and $\tilde j$, which correspond to sites $i$ and $j$ in the image, we set the blocks of $H_{pt}$ between $\tilde i$ and $\tilde j$ by
\begin{eqnarray}
\label{pullback}
{\rm dist}(\tilde i,\tilde j) \leq c_{inj}L & \; \rightarrow \; &
(H_{pt})_{\tilde i,\tilde j}=H_{f(\tilde i),f(\tilde j)} \\ \nonumber
{\rm dist}(\tilde i,\tilde j) > c_{inj}L & \; \rightarrow \; &
(H_{pt})_{\tilde i,\tilde j}=0.
\end{eqnarray}

The pullback Hamiltonian still obeys a locality bound, similar to Eq.~(\ref{decay}).  It is
\be
\label{decay2}
\sum_{j, {\rm dist}(i,j)\geq r} \Vert (H_{pt})_{ij} \Vert \leq J \exp(-Cr/R).
\ee

The Hamiltonian$H_{pt}$ need not have a gap due to the puncture; however, we will show use the next lemma to show Eq.~(\ref{bulkgap}) below which implies that
 $H_{pt}$ still has a gap ``in the bulk" away from the puncture in that for any vector $\phi$ supported sufficiently far from the puncture with $|\phi|=1$, $|H_{pt} \phi|$ is bounded away from zero. 
This lemma will also be useful later when we unfurl and will also be useful in theorem (\ref{mainFFlemma}).
\begin{lemma}
\label{localfn}
Consider a free fermion Hamiltonian $H$ on $T^d$ obeying
\be
\label{decayagain}
\sum_{j, {\rm dist}(i,j)\geq r} \Vert H_{ij} \Vert \leq J \exp(-r/R),
\ee
for all $i$.
Let $\phi$ be some state such that
\be
|(H-E) \phi| = \delta,
\ee
for some $E$; that is, $\phi$ is an approximate eigenvector of $H$.
Then, for any $\ell$, there is some sphere of radius $\ell$ such that there is a vector $\psi$ with $|\psi|=1$ and with $\psi$ supported on the intersection of that sphere with the support of $\phi$ such that
\be
\label{imply}
| (H-E) \psi| \leq \delta+O(J R/\ell).
\ee
\begin{proof}
Define a new Hamiltonian $H'$ such that $H'_{ij}=H_{ij}$ for ${\rm dist}(i,j)\leq \ell$ and $H'_{ij}=0$ for ${\rm dist}(i,j)>\ell$.  We will show the existence of a vector $\psi$ such that $|(H'-E) \psi| \leq \delta +O(JR/\ell)$ which will imply Eq.~(\ref{imply}) since $\Vert H'-H \Vert$ is exponentially small in $\ell/R$.  Note that $|(H'-E)\phi|=\delta'$ for $\delta' $ bounded by $\delta$ plus a quantity exponentially small in  $\ell/R$.

Let the torus have linear size $L$.  Without loss of generality, assume $\ell<L$.
Pick a random point $x$ and consider a sphere of linear size $\ell$ centered at that point.  Let $f(i)$ be a map from sites $i$ to reals given by
$f(i)=1-{\rm dist}(x,i)/\ell$ for ${\rm dist}(x,i)\leq \ell$ and $f(i)=0$ otherwise.
Let $\hat f$ be the block-diagonal matrix with $\hat f_{ii}=f(i) I$ where $I$ is the identity matrix in a block.

Let $\psi(x)=\hat f \phi$.  The probability that any given site is contained in that sphere is proportional to $(\ell/L)^d$, and the average of $|f(i)|^2$ over the sphere is of order unity, so the expectation value of $|\psi(x)|$ equals $c(\ell/L)^{d/2}$ for some constant $c$.  We will bound the expectation value of $|(H'-E)\psi(x)|$ by $c(\delta'+O(JR/\ell))(\ell/L)^{d/2}$.  Thus, the ratio of the expectation value of $|(H'-E)\psi(x)|$ to that of $|\psi(x)|$ is bounded by $\delta'+O(JR/\ell)$ so there is some choice of $x$ such that setting $\psi=\psi(x)$ obeys $|(H'-E) \psi| \leq \delta' +O(JR/\ell)$.

Note that
$(H'-E)\psi(x)=\hat f (H'-E) \phi + [(H'-E),\hat f] \phi$.
The expectation value of $|\hat f (H'-E) \phi|$ is proportional to $(\ell/L)^{d/2} \delta'$.
Let $P_n$ project onto the set of sites within distance $n\ell$ of $x$ for integer $n$.  Then,
$[(H'-E),\hat f] \phi=[(H'-E),\hat f] P_2 \phi+[(H'-E),\hat f] (1-P_2) \phi$.
Since $\hat f P_n=\hat f$ for $n>0$ and $H'$ vanishes between sites at least distance $\ell$ apart, $[(H'-E),\hat f] (1-P_2) \phi=0$.
So,
\be
| [(H'-E),\hat f] \phi | \leq |[(H'-E),\hat f] P_2 \phi|.
\ee

The expectation value of $|[(H'-E),\hat f] P_2 \phi|$ is bounded by $\Vert [(H'-E),\hat f] \Vert \cdot | P_2 \phi|$.   The expectation value of the norm $|P_2 \phi|$ is bounded by
$O(\ell/L)^{d/2}$.  The norm $\Vert [(H'-E),\hat f] \Vert$ is bounded by $O(J R/\ell)$.  To see this, Eq.~(\ref{decayagain}) implies that
$\sum_{j, mR\leq {\rm dist}(i,j)<(m+1)r} \Vert H_{ij} \Vert \leq J \exp(-m)$.
So,
$\sum_{j, mR\leq {\rm dist}(i,j)<(m+1)r} \Vert \Bigl( [(H'-E),\hat f] \Bigr)_{ij} \Vert \leq J (m+1) \frac{R}{\ell} \exp(-m)$,
and summing over $m=0,1,2,...$ gives $\Vert [(H'-E),\hat f] \Vert \leq O(J R/\ell)$.
\end{proof}
\end{lemma}

We now use lemma \ref{localfn} to show that for any vector $\phi$ with $|\phi|=1$ and with $\phi$ supported on sites on the punctured torus with distance at least $r$ from the puncture, we have that
\be
\label{bulkgap}
|H_{pt} \phi| \geq \Delta E - J O(\exp(-Cr/R))- O(JR/c_{inj}L).
\ee
To see this, apply lemma \ref{localfn} to $H_{pt}$ with $E=0$ and $\delta=|H_{pt}\phi|$.  Then, there is a vector $\psi$ supported on a sphere of radius
 $\ell$ with $|\psi|=1$ so that 
\be
\label{temp}
|H_{pt}\psi| \leq |H_{pt} \phi|+O(JR/\ell).
\ee
We
 pick $\ell=c_{inj}L$ so that the immersion is injective on that sphere.
The immersion maps $\psi$ to a state on the original system on  $R^d$.  To define this state, which we call $f(\psi)$, for $\tilde i$ in the sphere, we set $f(\psi)_{f(\tilde i)}=\psi_{\tilde i}$ while all other $f(\psi)_j$ are equal to $0$, where subscripts such as $\psi_{\tilde i}$ denote amplitudes of a vector in the subspace associated with a given site.
Note that $|H f(\psi)| \geq \Delta E$.
It is not necessarily the case that $|H_{pt}\psi|=|H f(\psi)|$, because in $H_{pt}$ we have removed terms in $H$ that connect sites inside the image of the immersion to those outside the image; however, we can bound the norm of such terms by $J O(\exp(-Cr/R))$ so $|H_{pt} \psi | \geq \Delta E-J O(\exp(-Cr/R))$ and combined with Eq.~(\ref{temp}) this gives Eq.~(\ref{bulkgap}).

If $L$ is at least a constant factor larger than $RJ/\Delta E$,
we can pick an $r$ of order $R J/\Delta E$ to make Eq.~(\ref{bulkgap}) give a nontrivial bound of at least $(1/2) \Delta E$ (any other constant smaller than unity multiplying $\Delta E$ would work as well; we pick $1/2$ for simplicity).  
Thus, we can modify the Hamiltonian $H_{pt}$ on sites within a distance $r$ from the puncture to give a new Hamiltonian which has a gap which is at least some constant times $\Delta E$.
 For large enough $L$ compared to $r$, these sites within distance $r$ of the puncture are not in $f^{-1}_{z,L}(Z)$.
We refer to this as ``healing the puncture".  Let the Hamiltonian that results from adding these terms be called $H'_{pt}$.

In appendix \ref{heal} we give an explicit construction that shows how to heal the puncture.

By doing this, we have worsened the locality properties of the Hamiltonian near the puncture, as the terms added near the puncture connect sites up to distance $r$, which may be a factor $J/\Delta E$ times larger than $R$.  To improve the locality properties we
define another map $g$ from $T^d$ to $T^d$; this map will map all sites within distance $r$ of the puncture to a single point and it will be a constant map for sites in $f^{-1}_{z,L}(Z)$ and it will obey
\be
{\rm dist}(g(x),g(y)) \leq C' {\rm dist}(x,y),
\ee
for some constant $C'$ of order unity.  Then, we use the function $g$ to ``pushforward" the Hamiltonian $H'_{pt}$ (that is, we just move where the sites are in $T^d$ according to the map $g$, without changing the Hamiltonian).  This gives a new Hamiltonian that we call $H'_{torus}(H,Z)$, that fulfills the claims of theorem \ref{fftorus}.

The Hamiltonian $H'_{torus}(H,Z)$ is then a gapped Hamiltonian on the torus.  We can unfurl this Hamiltonian to a Hamiltonian on the whole $R^d$.  This is done by defining a covering map from $R^d$ to $T^d$ and using this map to pull back the Hamiltonian $H'_{torus}(H,Z)$ to a Hamiltonian $H'(H,Z)$.
This pullback is defined similarly to our definition of the pullback of a Hamiltonian when we constructed the immersion:
given two sites in the pre-image of the covering map, called $\tilde i$ and $\tilde j$, which correspond to sites $i$ and $j$ in the image of the covering map, we set the matrix element of $H_{pt}$ between $\tilde i$ and $\tilde j$ by
\begin{eqnarray}
{\rm dist}(\tilde i,\tilde j) \leq L/4 &\; \rightarrow \; & 
(H'(H,Z))_{\tilde i,\tilde j}=(H'_{torus}(H,Z))_{f(\tilde i),f(\tilde j)}, \\ \nonumber
{\rm dist}(\tilde i,\tilde j)  > L/4 &\; \rightarrow \; & 
(H'(H,Z))_{\tilde i,\tilde j}=0.
\end{eqnarray}
 The  distance $L/4$ is chosen to be some quantity small enough compared to $L$ that the covering map is injective on distances smaller than this.
This is a general principle in constructing a pullback of a Hamiltonian: the interaction terms must become small at the length scale at which the map becomes non-injective.

For sufficiently large $L$, we can show that this Hamiltonian $H'(H,Z)$ has a gap using lemma \ref{localfn}; there are some technical details needed to do this as what we must do is consider normalized states $\phi$  in the infinite system in $R^d$; then, since $\phi$ is normalized, we can restrict it to a finite region with size small compared to $L$ with only a small change in $|H \phi|$; then we embed this finite region in a torus and apply the lemma to show that if there is a state $\phi$ with $|H\phi|$ small compared to  $\Delta E$, then there is a state $\psi$ supported on a sphere of radius small compared to $L$ with $|H\psi|$ small compared to $\Delta E$.  However, the gap in $H'_{torus}$ implies that no such $\psi$ exists.

Note that we can map Hamiltonian $H'_{torus}(H,Z)$ to a Hamiltonian on a smaller torus of size $L'$ with $L<L' < 2\pi L$ by an injective map that leaves distances and angles invariant in the hypercube whose image under the immersion is $Z$.  Using this map, we can change the periodicity of $H'(H,Z)$ as was mentioned below theorem \ref{fftorus}.

\subsection{Classifying Periodic and Aperiodic Systems}
We now combine the above result with a classification of periodic systems to classify aperiodic systems.  We begin by reviewing the case of periodic Hamiltonians.
Given a periodic Hamiltonian, we compute its bandstructure.  Since we will consider periodic Hamiltonians obtained by unfurling a torus, the Brillouin zone is also a torus which we parameterize by angles $\theta_1,...,\theta_d$.  The bandstructure defines a Hamiltonian $H(\theta_1,...,\theta_d)$, which depends smoothly on the angles, with the dimension of the Hamiltonian being equal to the number of sites in a unit cell.  Conversely, given any Hamiltonian $H(\theta_1,...,\theta_d)$ which depends smoothly upon angles, we can construct a periodic Hamiltonian whose band structure is precisely $H(\theta_1,...,\theta_d)$ by an inverse Fourier transform; then, the smooth dependence of $H(\theta_1,...,\theta_d)$ upon angles implies a rapid decay of matrix elements in space.  So, we use the terms ``periodic Hamiltonian" and $H(\theta_1,...,\theta_d)$ interchangeably.

Given such a Hamiltonian, and assuming that there is a gap in the spectrum near zero for all points in the Brillouin zone, we can define a projector $P(\theta_1,...,\theta_d)$ onto negative energy states which also depends smoothly on the angles.  Throughout, when we discuss any operator depending upon angles, we will assume it is smooth (infinitely differentiable). 
A projector $P(\theta_1,..,\theta_d)$ defines a vector bundle.  These vector bundles are classified by K-theory classes, which do not change under continuous deformations.
We briefly review the fact that that given two gapped periodic Hamiltonians $H_0(\theta_1,...,\theta_d)$ and $H_1(\theta_1,...,\theta_d)$ we can stabilize (add additional sites to the unit cells of $H_0$ and $H_1$ with no matrix elements connecting those sites to other sites) and then connect the Hamiltonians by a continuous path of gapped periodic Hamiltonians $H_s(\theta_1,...,\theta_d)$ if and only if the
K-theory class is the same.
For the ``only if" direction of this, note that 
 given a family of periodic Hamiltonians which depend continuously on a parameter $s$, we can define a continuous family of projectors $P_s(\theta_1,...,\theta_d)$ and the K-theory class does not change under continuous deformation.

For the ``if" direction of this, suppose that $P_0(\theta_1,...,\theta_d)$ and $P_1(\theta_1,...,\theta_d)$ are in the same K-theory class, so by stabilizing (direct sum with a projector that does not depend upon $\theta$) we can connect them by a continuous path of projectors which also depend smoothly upon angles.  This stabilization (by direct sum with a projector that does not depend upon $\theta$) can be obtained precisely by adding additional sites to $H_0$ and $H_1$ with no matrix elements connecting those sites to others.  
So, we add those sites.  After adding these sites, we ``spectrally flatten"; that is, we find a continuous path of Hamiltonians from $H_0(\theta_1,...,\theta_d)$ to $J(1-2P_0(\theta_1,...,\theta_d))$.  Then follow a continuous path from $J(1-2P_0(\theta_1,...,\theta_d))$ to $J(1-2P_1(\theta_1,...\theta_d))$ and finally use the spectral flattening of $H_1(\theta_1,...,\theta_d)$ to construct a continuous path from $J(1-2P_1(\theta_1,...,\theta_d))$ to $H_1(\theta_1,...\theta_d)$.  This gives a continuous path from $H_0$ to $H_1$.
The spectral flattening can be constructed as follows: let $H_0$ have a spectral gap $\Delta E$ near energy $0$.  Define a family of functions $f_t(x)$ which depends continuously on $t$ and smoothly on $x$, with $f_0(x)=x$ and $f_1(x)=J$ for $x \leq -\Delta E$ and $f_1(x)=0$ for $x\geq \Delta E$.  Then, let $H_{0,t}(\theta_1,...,\theta_d)=f_t(H_0(\theta_1,...,\theta_d))$.

Note that the periodic Hamiltonians might have an interaction range larger than a single unit cell.  In contrast, the torus trick above constructs a periodic Hamiltonian with a unit cell size larger than the interaction range $R$.

Certain tricks will be re-used several times in what follows so we mention them briefly here and explain in more detail below.  Recall that we refer to block-diagonal gapped Hamiltonians as trivial.  We will also call any Hamiltonian trivial if it can be deformed to such a Hamiltonian.  One trick is 
that for any gapped Hamiltonian $H$, the direct sum $H \oplus -H$ is trivial.  A second fact is that while we have considered paths where we deform Hamiltonians, we could instead keep the terms in the Hamiltonian fixed and deform where the sites are in the ambient space.  We could have allowed this deformation of where the sites are as part of the definition of a path of Hamiltonians but this is not necessary if we consider stable equivalence as if $H$ and $H'$ are two Hamiltonians that differ only in a slight displacement of the sites, then
$H \oplus H' \oplus -H'$ is trivial, and $H \oplus -H'$ can be deformed to a diagonal Hamiltonian in a similar way to how $H \oplus -H$ can in Eq.~(\ref{twobytwo}), so $H$ can be deformed into $H'$ up to stable equivalence.

We now claim that:
\begin{theorem}
\label{mainFFlemma}
Consider any two free fermion Hamiltonians $H_0,H_1$ whose interactions decay following Eq.~(\ref{decay}) with given $J,R$ and which both have gap at least $\Delta E$, and consider any two disjoint hypercubes $Z_0,Z_1$ with linear size $L$ with $L$ sufficiently large compared to $R J/\Delta E$.

{\bf 1.} If $H'_{torus}(H_0,Z_0)$ is in the same K-theory class as $H'_{torus}(H_1,Z_1)$, there exists a Hamiltonian $H$ which obeys Eq.~(\ref{decay}) with the same $J$ and which has range that is upper bounded by a constant times $R$ and which has a gap which is at least lower bounded by a positive constant times $\Delta E$ such that Hamiltonian $H$ agrees with $H_0$ on $Z_0$ and also agrees with $H_1$ on $Z_1$.  We say that such a Hamiltonian $H$ ``interpolates between $H_0$ on $Z_0$ and $H_1$ on $Z_1$".

{\bf 2.} If $H'_{torus}(H_0,Z_0)$ is not in the same K-theory class as $H'_{torus}(H_1,Z_1)$, then there is no Hamiltonian $H$ which obeys Eq.~(\ref{decay}) with the given $R,J,\Delta E$ which interpolates between $H_0$ on $Z_0$ and $H_1$ on $Z_1$.
\end{theorem}

Before giving the proof, we make two remarks.  First, if a Hamiltonian has a gap $\Delta E$, then it has a gap at least $\Delta E'$ for any $\Delta E'<\Delta E$, and similarly if it obeys Eq.~(\ref{decay}) for any given $R,J$ it also obeys that equation for any larger $R',J'$.  This is useful in applying the second claim {\bf 2}; suppose we have two Hamiltonians $H_0,H_1$ with given $R,J,\Delta E$ and we wish to show that there is no Hamiltonian with, for example, a range $10 R$ and a gap $\Delta E/10$ that agrees with $H_0$ on one hypercube and $H_1$ on another hypercube.  To do this, we regard our initial Hamiltonians as obeying Eq.~(\ref{decay}) with range $10R$ and gap $\Delta E/10$ and then apply statement {\bf 2}.  We find then that we need to have $L$ be sufficiently large compared to $100 R J/\Delta E$.  The second remark is that when we consider K-theory classes for bundles over a torus, there are often ``lower dimensional invariants".  For example, bundles over $T^3$ without symmetry are classified by $Z \oplus Z \oplus Z$, but all three of these invariants are in a sense ``lower dimensional", and are obtained by considering the dependence upon only two of the angles at a time.  The map from $H$ to $H'_{torus}(H,Z)$ only produces Hamiltonians with trivial values of these lower dimensional invariants even if $H$ itself is periodic and has nontrivial values of these invariants.  This resolves an apparent paradox: claim {\bf 2} implies that we cannot find a gapped Hamiltonian that interpolates spatially between two $H'$ with different values of these invariants, but we know that in fact we can interpolate between two periodic Hamiltonians with different values of these invariants.  That is, the resolution of the apparent paradox is that $H'_{torus}$ ``forgets" some of the invariants of a periodic Hamiltonian $H$.

\begin{proof}
We begin with the first claim.  First we construct a Hamiltonian $h$ that agrees with $H_0$ on $Z_0$ and so that on some other hypercube $X_1$ it agrees with $H_1$ on $Z_1$, where $X_1$ and $Z_1$ may be related by a translation and rotation.  We will in turn describe the construction of that Hamiltonian in two steps; first we will construct a Hamiltonian that does the desired interpolation but does not satisfy useful bounds on its range and gap and then we will correct the problem by constructing another Hamiltonian that does have useful bounds on range and gap.
To do the first step, consider the torus that $H'_{torus}(H_0,Z_0)$ is defined on, and unfurl this torus to $R^d$, giving a tiling of $R^d$ with hypercubes.  Then define a Hamiltonian that interpolates between $H'_{torus}(H_0,Z_0)$ for hypercubes near the hypercube containing $Z_0$ to $H'_{torus}(H_1,Z_1)$ for hypercubes far away.  We use the existence of a path of periodic gapped Hamiltonians that connects $H'(H_0,Z_0)$ to $H'(H_1,Z_1)$ to define this interpolation.
If the interpolation is done sufficiently slowly over space, then lemma (\ref{localfn}) allows us to show a lower bound on the gap: if there is a state $\psi$ such that this interpolating Hamiltonian acting on that state is small, then by lemma (\ref{localfn}) there is a state on a region of size of order $RJ/\Delta E$ such that the interpolating Hamiltonian acting on that state is small, and then we can apply the gap of the interpolating Hamiltonians.

This interpolating Hamiltonian construction has one problem as we mentioned: the interpolating periodic Hamiltonians may have a range that is much larger than $R$.
To solve this problem, first we zero all terms in the interpolating Hamiltonian connecting sites $i,j$ with large ${\rm dist}(i,j)$; if we only do this for sufficiently large distance, then this does not close the gap.
We pick a quantity $R_1$ of order $L$ so that the entire hypercube $Z$ is contained in a sphere of radius $R_1$ center at a point $z_0 \in Z_0$.
Choose the interpolating Hamiltonian to agree with $H'(H_0,Z_0)$ up to some distance $R_2>>R_1$ from $Z_0$ and to agree with $H'(H_1,Z_1)$ beyond some distance $2R_2$ from $Z_0$.  Define a piecewise linear function $f(x)$ by $f(x)=x$ for $0 \leq x \leq R_1$ or $x \geq 4R_2$.  For $R_1 \leq x \leq 2R_2$, $f(x)=R_1$ and for $2R_2 \leq x \leq 4R_2$, $f(x)$ is a linear function interpolating between $f(2R_2)=R_1$ and $f(4R_2)=4R_2$.
 Define a map of $R^d$ to itself by mapping a point $z$ in at distance $r$ from $z_0$ to the point along the line from $z$ to $z_0$ which is at distance $f(r)$ from $z_0$.  This map is the identity both near $Z_0$ and also sufficiently far from $Z_0$.  Then use this map to pushforward the Hamiltonian that we have constructed.  For sites $i,j$ with $2R_1 \leq {\rm dist}(i,z_0),{\rm dist}(j,z_0) \leq R_2$,
the distance ${\rm dist}(i,j)$ is mapped to zero if $i,j$ are both along the same line from $z_0$, while if $i,j$ are not along the same line then the distance is reduced by a factor of at least $O(R_1/R_2)$.  So, by choosing $R_2/R_1$ large, this gives a Hamiltonian $h$ with bounded range as we have compressed all the interpolating Hamiltonians to the sphere at distance $R_1$ from $z_0$.  (In fact, this argument might lead to the $J$ in the Hamiltonian $h$ being a constant factor larger than the original $J$ depending on the norm of the interpolating Hamiltonians; however, we can then multiply $h$ by a scalar less than one to restore the original value of $J$ with only a constant factor reduction in the gap).
Having constructed the desired $h$, we can apply a further map to move some hypercube $X_1$ at distance larger than $R_2$ from $z_0$ to the desired position $Z_1$, giving the desired interpolating Hamiltonian $H$.

Now we show the second claim.  Suppose such an interpolating Hamiltonian does exist.  Then, consider a path of hypercubes $Z_s$ that starts at $Z_0$ and ends at $Z_1$ by continuously sliding and translating the hypercube.  Such a path of hypercubes defines a path of periodic Hamiltonian $H'(H,Z_s)$.  This path of Hamiltonians may have discontinuities.  We will show that, for sufficiently large $L$ compared to $RJ/\Delta E$, the K-theory class does not change across these discontinuities, which implies that $H'(H,Z_0)$ and $H'(H,Z_1)$ are in the same K-theory class.  One source of discontinuities comes from Eq.~(\ref{pullback}): we set certain matrix elements to zero when the distance between two sites on the torus becomes sufficiently large.  However, since the matrix elements are exponentially small, if we replace the discontinuous jump in matrix elements by a continuous path, the gap does not close along this path so the K-theory class does not change.

A more important source of discontinuities comes from changes in how we heal the puncture.  These change the Hamiltonian in a region of size of order $R$ around the puncture but not elsewhere.  Importantly, this change happens on a set of size small compared to the size of the torus.  We now show that this does not change the K-theory class.
We need the following result: consider any Hamiltonian $H$.  Define a new system by doubling the Hilbert space dimension on each site. Then the Hamiltonian $H \oplus -H$ on this new system can be continuously deformed to a diagonal Hamiltonian while keeping the gap open and keeping the interaction range bounded.  The path can be given explicitly as
\be
\label{twobytwo}
\begin{pmatrix}
(1-s) H & s I \\ s I & -(1-s)H
\end{pmatrix}.
\ee
The off-diagonal elements of this block matrix are proportional to the identity so now the Hamiltonian is block-diagonal.  By diagonalizing $H$ we see that for every eigenvalue $E$ of $H$, the Hamiltonian in the path above has eigenvalues given by the eigenvalues of the two-by-two Hamiltonian
\be
\label{2b22}
\begin{pmatrix} (1-s) E & s \\ s & -(1-s) E \end{pmatrix},
\ee
and so the gap remains open.
For classes with sublattices symmetry, we maintain the symmetry in the above path by declaring the sublattices to be interchanged in the system $-H$ compared to that in $+H$.
At the end of this path, the Hamiltonian is block-diagonal.

A similar path can be used to deform $H \oplus -H'$ to a diagonal Hamiltonian if $H$ and $H'$ are two Hamiltonians that differ only in that the sites of $H'$ are slightly displaced from those of $H$.  However, then at the end of the path, the Hamiltonian is block-diagonal with each block corresponds to a pair of sites.  Such a Hamiltonian can then be easily deformed to a block-diagonal Hamiltonian with each block corresponding to a single site.  In symmetry classes other than class A, we need to use the requirement, made at the start of the section, that symmetry operations for time-reversal and particle-hole symmetry are block-diagonal and that for sublattice symmetry the subspace on each site is the direct sum of two subspaces of the same dimension, corresponding to the two different sublattices.

Now, consider two Hamiltonians $H_0$ and $H_1$ that agree everywhere except within distance $R$ of the puncture.  Consider the Hamiltonian $H_0 \oplus H_1 \oplus -H_1$.  Using the above path, $H_1 \oplus -H_1$ is equivalent to a trivial Hamiltonian, so $H_0 \oplus H_1 \oplus -H_1$ is stably equivalent to $H_0$.  Now we construct a path of Hamiltonians $H_s$ that agrees with $H_0 \oplus -H_1$ near the puncture and agrees with
\be
\begin{pmatrix}
(1-s) H_0 & s I \\ s I & -(1-s)H_1
\end{pmatrix}.
\ee
far from the puncture.  One explicit way to do this is to define a function $r_i$ that is $0$ for sites near the puncture, $1$ for sites far from the puncture, and interpolates between.  Promote this function to an operator $\hat r$ which is a block-diagonal matrix with $(\hat r)_{ii}=r_i I$.  Then consider the path of Hamiltonians
\be
\label{Jspath}
J_s=
\begin{pmatrix}
\sqrt{1-s \hat r} H_0 \sqrt{1-s \hat r} & s \hat r \\ s \hat r & -\sqrt{1-s \hat r} H_1 \sqrt{1-s \hat r}
\end{pmatrix}.
\ee
Again using lemma \ref{localfn} we can show that this path of Hamiltonians has a gap.  $J_1$ has nonvanishing offdiagonal matrix elements only near the puncture (i.e., only on a contractible set) so we can deform it to a block diagonal Hamiltonian so it has trivial K-theory class.  So, $H_0 \oplus - H_1$ has trivial K-theory class and since $H_0$ is stably equivalent to  $H_0 \oplus H_1 \oplus -H_1$, it implies that $H_0$ and $H_1$ have the same K-theory class.
\end{proof}
We will use the procedure in Eq.~(\ref{Jspath}) again later, so we give it a name: we say that we ``join $H_0$ and $-H_1$ far from the puncture".

\begin{figure}
\includegraphics[width=3in]{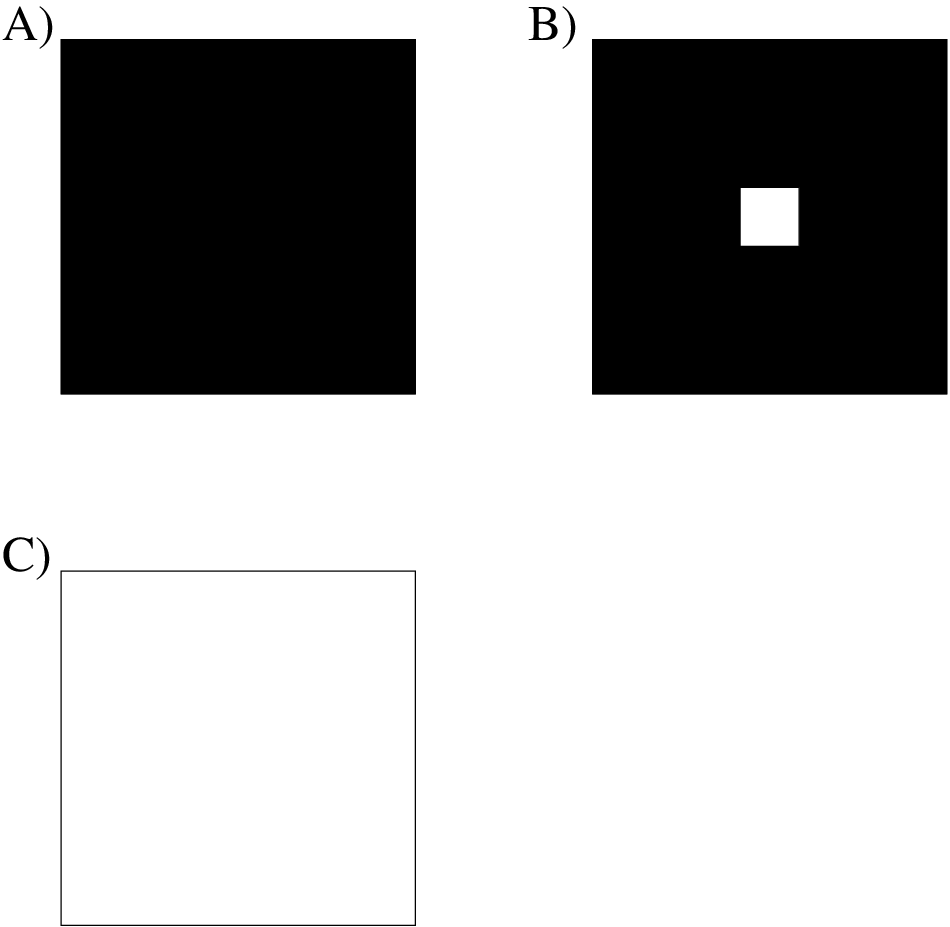}
\caption{a)Initial system b)Creating hole in system c)Stretching to a lower dimensional system}
\label{FigStretch}
\end{figure}

Using this result on the existence of interpolating Hamiltonians we can also show that
\begin{theorem}
\label{pathFFlemma}
Let $H_0$ and $H_1$ be Hamiltonians with ambient space $M=T^d$ obeying Eq.~(\ref{decay}) with decay constant $R$ and with gap $\Delta E$, such that $H'(H_0,Z)$ and $H'(H_1,Z)$ are in the same K-theory class for some hypercube $Z$ with linear size $L$ sufficiently
large compared to $RJ/\Delta E$.
Let $S$ be any contractible set.

Then, up to stable equivalence
there exists a continuous path of Hamiltonians from $H_0$ to some Hamiltonian which agrees with $H_1$ on $S$, with the Hamiltonians in this path
having range bounded by a constant times $R$ and gap lower bounded by a positive constant times $\Delta E$.
\end{theorem}
Note that we only show that we can find a path to a Hamiltonian which agrees, up to stable equivalence, with $H_1$ on a contractible set.  In fact, even if the conditions of the theorem hold, there may be an obstruction due to lower-dimensional invariants to finding a path from $H_0$ to $H_1$ itself.

\begin{figure}
\includegraphics[width=3in]{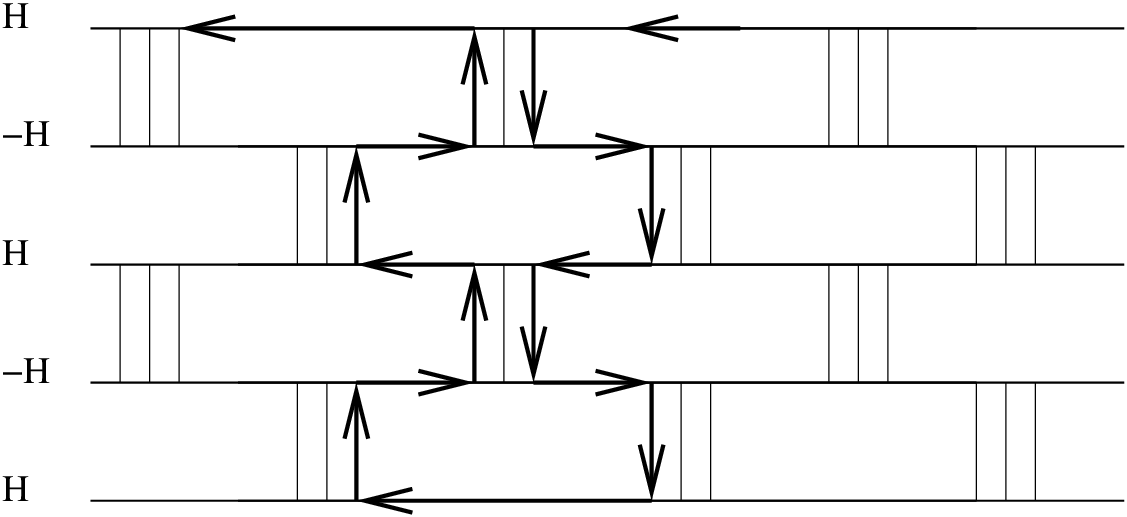}
\caption{Joining $H \oplus -H \oplus H \oplus -H \oplus H$.  Vertical lines denote join.  Arrow sketches part of remaining system after removing sites inside join.}
\label{FigMakeMore}
\end{figure}

\begin{proof}
We will show that if $H'(H_0,Z)$ is in the trivial K-theory class then we can deform $H_0$ to a Hamiltonian which is block-diagonal on $S$.  This will imply the theorem, as follows: let $H_0,H_1$ be such that the conditions of the theorem hold.  Then $H_0 \oplus -H_1$ can be deformed to a Hamiltonian which is block-diagonal on $S$ (we use the fact that $-H_1$ has the opposite K-theory class to $H_1$).
So $H_0 \oplus -H_1 \oplus H_1$ can be deformed to a Hamilton which agrees with $H_1$ on $S$ up to stable equivalence; however, since $-H_1 \oplus H_1$ can be deformed to a block-diagonal Hamiltonian,
$H_0 \oplus H_1 \oplus -H_1$ is stably equivalent to $H_0$, and so $H_0$ can be deformed to a Hamiltonian which agrees with $H_1$ on $S$ up to stable equivalence.

Fig.~\ref{FigStretch} shows the steps of the path that we construct.  We illustrate the path in the case $d=2$, while calculations in other dimensions are done similarly.
The torus is drawn by identifying boundaries of a square and we will illustrate the path for the case that $S$ is chosen to include all points except those on the boundary of the square.

Fig.~\ref{FigStretch}A shows the initial system in black.   Then, we find a path to make the Hamiltonian diagonal on a square of linear size of order $L$; this square is on shown in white in Fig.~\ref{FigStretch}B.  To make the Hamiltonian diagonal on a square, we do the following steps: {\bf 1} construct a Hamiltonian $O$ which agrees with $H_0$ on the given square and becomes block-diagonal sufficiently far from the square, with appropriate bounds on the range and gap of $O$.  We can construct such an $O$ using a construction similar to that of theorem \ref{mainFFlemma}: since $H'(H_0,Z)$ is in the trivial K-theory class, we can find a Hamiltonian that interpolates from $H'(H_0,Z)$ to a block-diagonal Hamiltonian which also is in the trivial K-theory class.  {\bf 2} Consider $H_0 \oplus O \oplus -O$.  This is stably equivalent to $H_0$.  Also, up to stable equivalence we can remove those sites far from the square where $O$ and $-O$ are block-diagonal.  {\bf 3} Having removed the sites, move the remaining sites in $O$, contracting them to a single point, and move that point to somewhere in the black region.  Up to stable equivalence we can now remove that point, leaving Hamiltonian $H_0 \oplus -O$.  {\bf 4} The next step is to join $H_0$ and $-O$ near the square.  This step is similar to Eq.~(\ref{Jspath}) except we now define the function
$r_i$ to be $1$ for sites {\it in} the square and $1$ for sites {\it far} from the square and interpolating between, and we consider the path of Hamiltonians
\be
\begin{pmatrix}
\sqrt{1-s \hat r} H_0 \sqrt{1-s \hat r} & s J \hat r \\ s J \hat r & -\sqrt{1-s \hat r} -O \sqrt{1-s \hat r}
\end{pmatrix}.
\ee
This path keeps the gap open and leaves the Hamiltonian block-diagonal in the square at the end), and so then we can remove the sites in the square.

Given the configuration in Fig.~\ref{FigStretch}B, we would like to ``stretch" the system, moving the sites so that they are as in Fig.~\ref{FigStretch}C.  However, moving them like this violates the assumption on decay with range $R$.  So, first we modify before stretching.  Consider first a one-dimensional system.  Up to stable equivalence, we can deform $H$ to $H \oplus (-H \oplus H) \oplus (-H \oplus H) \oplus ...$, adding $n$ copies of $-H \oplus H$ for some given $n$.  Then, we follow a procedure of joining copies of $H$ to copies of $-H$ as shown in Fig.~\ref{FigMakeMore}.  Each horizontal line denotes a copy of $H$ or $-H$ (we have shown the case $n=2$).  The vertical lines denote places where we join one copy to another.  The horizontal length of each joined region is chosen to be large compared to $RJ/\Delta E$ so that the join keeps the gap open and the length between joined regions is also large compared to $RJ/\Delta E$.  We drop all terms in the Hamiltonian that couples sites on the left side of a join to sites on the right side of a join (for sufficiently large joins, this does not close the gap).  In the middle of the join, the two copies of the Hamiltonian which are joined become diagonal and so sites in the middle of the join can be remove.  After removing the sites in the middle of the join, we have sketched part of the remaining system using a thickened line with an arrow (to avoid cluttering the figure, we have sketched only part of the remaining system; the thickened line with an arrow should continue all the way around the figure).  Next, we move the sites in the remaining system; the length of the arrow is roughly $n\Ls$, and so we can map the remaining system onto a torus of size $\Ls$ by a map that reduces distance by a factor $O(1/n)$ so that the interaction range becomes $O(n/R)$.
We have described this procedure in one dimension, but it works equally well in any dimension; we simply perform the procedure for each of the $d$ different axes of the torus in turn, treating the system as if it were a $1$-dimensional system by ignoring the other $d-1$-dimensions.

After doing this, if we choose $n \sim \Ls/R$, we can stretch the system as in Fig.~\ref{FigStretch}C.
\end{proof}

\section{Quantum Cellular Automata and Local Unitaries}
Having discussed free systems, we next turn to interacting systems and QCA\cite{QCA}.  These QCA will play an important role in this paper, both for their own sake and for their application to interacting systems, so we take some time to review them here.  We will mention a distinction between QCA and quantum circuits highlighted in Ref.~\onlinecite{QCA}, and we will draw an analogous distinction between what we will call {\em locally-generated unitaries} (LGU) and {\em locality-preserving unitaries} (LPU).  This distinction between the later two possibilities seems not to have been highlighted before.

\subsection{Quantum Circuits and Quantum Cellular Automata}
Consider a quantum system on a lattice in $d$ dimensions.  On each site, there is a finite dimensional Hilbert space.  If the lattice is finite, then defining a quantum circuit is simple: it is a unitary $U_{QC}$ such that $U_{QC}$ can be written as the composition of several unitaries $U_i$:
\be
U_{QC}=U_r U_{r-1} ... U_1,
\ee
where $r$ is some integer labeling the number of ``rounds" of the quantum circuit, and where each $U_i$ is a unitary with the property that 
$U_i$ is a product of unitaries $U_{i,X}$ supported on disjoint sets $X$ with some bound $R_i$ on the diameter of each set $X$.  Each such $U_{i,X}$ is referred to as a ``gate".  We let $S_i$ denote the set of sets $X$ such that there is a gate $U_{i,X}$ in the $i$-th round.  We refer to $R_i$ as the ``range" of the $i$-th round and let $\sum_{i=1}^r R_i$ be the range of the quantum circuit.

A QCA is defined to be a unitary transformation $U$ such that for any set of sites $A$ and any operator $O_A$ supported on some set $A$, then $U O_A U^\dagger$ is supported on the sites of sites within distance $R$ of $A$, for some given $R$.  We refer to $R$ as the range of the QCA.  Note that every quantum circuit with range $R$ is a QCA with range $R$.
For an operator $O_A$ supported on a set $A$, the operator $U_i O_A U_i^\dagger$ is supported on the set $A \cup_{X \in S_i, X\cap A \neq \emptyset} X$.  Hence given a bound on $r$ and on the diameter of the sets $X$, it follows that for any operator $O_A$ supported on some set $A$, the operator $U_{QC} O_A U_{QC}$ is supported within a bounded distance of $A$; one can think of this as a ``light cone", with the support of the operator increasing from one round to the next.

If the lattice is infinite, then a more complicated definition is necessary.  First, on every finite set one defines an algebra of operators supported on that set.
Next, one defines a closure of this algebra on an increasing family of sets; this closure is called the {\em quasi-local algebra}.  
Then, one defines a QCA as an automorphism $\alpha_{QCA}$ of this quasi-local algebra, such that that for any set of sites $A$ and any operator $O_A$ supported on some set $A$, 
$\alpha_{QCA}(O_A)$ is supported on the set of sites within distance $R$ of $A$.

Note the bound on the range of $U_{QCA}$ or $\alpha_{QCA}$ implies the same bound $R$ on the range of the inverse $U^\dagger_{QCA}$ or $\alpha^{-1}(QCA)$.  To see this, consider an operator $O_i$ supported on a site $i$.  We wish to show that $\alpha^{-1}(O_i)$ is supported on sites within distance $R$ of $i$; however, this is equivalent to showing that $[\alpha^{-1}(O_i),O]=0$ for all operators $O$ supported more than distance $R$ from site $i$.  However, $[\alpha^{-1}(O_i),O]=\alpha^{-1}([O_i,\alpha(O)])$, and using the bound on the range of $R$ we have that $[O_i,\alpha(O)]=0$.

Note that for a unitary $U$ on a finite system, the map $O \rightarrow U O U^\dagger$ is an automorphism of the algebra of operators; conversely, such an algebra  automorphism on a finite system is always of the form $O \rightarrow U O U^\dagger$ and the automorphism determines $U$ up to a scalar.  So, we will sometimes choose to refer to a QCA for a finite systems as an automorphism $\alpha_{QCA}$ rather than as a unitary $U_{QCA}$ if we do not care about the phase.

One defines a quantum circuit for an infinite system as an automorphism $\alpha_{QCA}$ which is the composition of $r$ automorphisms $\alpha_i$ of the quasi-local algebra, such that: for all $i$ there is a set $S_i$ of disjoint sets with bounded diameter such that $\alpha_i$ is an automorphism of the algebra of observables on $X$ for all $X$ in $S_i$ and such that $\alpha_i$ is the identity map on the algebra of observables supported on sites not in any $X$ for $X \in S_i$. 
Alternately, a useful definition for infinite systems is to consider families of QCAs, defined on increasing size systems, whose actions on operators supported on any fixed set converge to some limit, with uniform bounds on the range of the QCAs.

While every quantum circuit is a QCA, the converse is not true.  In Ref.~\onlinecite{QCA}, it was shown that for infinite one-dimensional systems, QCA without any symmetry properties are classified by an index which is a positive rational, as we review in subsection \ref{inoned}.  
A similar result holds for finite systems: one can define a family of QCAs on finite systems of increasing size, so that the QCA has fixed range $R$ but so that the smallest range of a quantum circuit realizing the given QCA diverges system size.

\subsection{Locally Generated Unitaries}
The definitions of quantum circuits and QCA are useful, but in many cases we would like to soften the definition somewhat, replacing the strict locality with a softer notion.  We refer to the resulting concepts as locally generated unitaries (LGU) and locality-preserving unitaries (LPU).

For a finite system, we define an LGU to be the unitary generated by evolution for a fixed time under a time-dependent Hamiltonian whose interactions decay sufficiently rapidly (sufficiently fast polynomials will suffice for a finite dimensional system, but the most interesting case will be a superpolynomial decay).
More formally, we let
\be
U_t={\cal S} \, \exp[i\int_0^t H_s {\rm d}s],
\ee
where $H_s$ is a Hermitian matrix that depends upon a parameter $s$, and where ${\cal S}$ and $s$-ordered exponential.  For any fixed $t$, $U_t$ is an LGU.
We will impose a requirement that the interactions of $H_s$ decay rapidly with space.  Various possibilities have been considered and fall under the general term of ``Lieb-Robinson bounds"\cite{lr1,lsm,lr2,lr3,lr4}, and we just list one possibility in the next paragraph.  The key result is the bound Eq.~(\ref{LRbound}) below.

Let
\be
H_s=\sum_{i} \sum_{R \geq 0} H_{i,R}(s),
\ee
where each $H_{i,R}(s)$ is supported on the set of sites within distance $R$ of site $i$.  We require the interactions to decay rapidly by
\be
\label{Kbound}
\sum_{R' \geq R} \Vert H_{i,R'}(s) \Vert \leq K(R),
\ee
for some function $K$.  We require that $K$ obey the following property, which we call ``reproducing":
\be
\sum_m K({\rm dist}(i,m)) K({\rm dist}(m,j)) \leq \lambda K({\rm dist}(i,j)),
\ee
for some constant $\lambda$, where the sum is over sites of the lattice.
For a square lattice in $d$ dimensions and the Euclidean metric, the powerlaw $K(l) \sim l^{-\alpha}$ is reproducing for sufficiently large $\alpha$.  An exponential decay is {\it not} reproducing, but an exponential multiplying a sufficiently fast powerlaw is.  Therefore, if Eq.~(\ref{Kbound}) holds for an exponentially decaying $K$ for some given decay rate, it holds for a slightly slower exponential decay rate multiplied by a fast decaying powerlaw.
For reproducing $K$, there is the bound that for any operator $O_A$ supported on set $A$ and any $O_B$ supported on set $B$,
\be
\label{bd1}
\Vert U_t O_A U_t^\dagger, B] \Vert \leq 2 \Vert O_A \Vert \Vert O_B \Vert \sum_{i \in A} K({\rm dist}(i,B)) [\exp(2 \lambda t)-1].
\ee
If $K$ is exponentially decaying, this enables us to define a ``Lieb-Robinson velocity" $v_{LR}$ such that for $|t| \leq {\rm dist}(A,B)/v_{LR}$ the commutator is exponentially small.  On the other hand, if $K$ decays slower than exponential (for example as $\exp(-t^\alpha)$ for some $0 < \alpha <1$), the bound is still effective for any fixed $t$ but there is no upper bound to the propagation speed.

Let $b_R(A)$ denote the set of sites within distance $R$ of a set $A$.
Given the bound (\ref{bd1}), it follows that for any operator $O_A$ supported on a set $A$, and for any $R$, there is an operator $O_{b_R(A)}$ supported on $b_R(A)$ such that
\begin{eqnarray}
\label{LRbound}
\Vert U_t O_A U_t^\dagger - O_{b_R(A)} \Vert   & \leq & 2C \Vert O_A \Vert \sum_{i \in A} K({\rm dist}(i,\overline{b_R(A)})) \\ \nonumber
& \leq & 2C \Vert O_A \Vert \sum_{i \in A} K(R+{\rm dist}(i,\overline A)),
\end{eqnarray}
for some constant $C=\exp(2 \lambda t)-1$, where $\overline A$ denotes the complement of $A$.
The expression in the second line,
$\sum_{i \in A} K(R+{\rm dist}(i,\overline A))$, can be expressed as $N_0 K(R+1) + N_1 K(R+2) + N_2 K(R+3) + ...$, where $N_0$ is number of points on the boundary of $A$, $N_1$ is the number of points in $A$ which are distance $1$ away from the boundary, and so on.

The theory of classifying different phases is closely related to that of LGUs.
Given a differentiable path of Hamiltonian $H_s$ with a gap and local interactions, then the technique of quasi-adiabatic continuation\cite{lsm,hastingswen,bravyihastings,bhm,tjo} allows us to define an LGU which maps the ground state of $H_0$ to that of $H_1$.
A fundamental role in this is play by the following elementary identity: if $U \psi_0=\psi_1$, where $\psi_0$ and $\psi_1$ are the ground state of $H_0$ and $H_1$ respectively, then
\be
\langle \psi_1 | O | \psi_1 \rangle = \langle \psi_0 | \Bigl( U^\dagger O U \Bigr) | \psi_0 \rangle.
\ee
If $O$ is a local operator then $\Bigl( U^\dagger O U \Bigr) $ can be approximated by an operator supported near the support of $O$ (that is, we use the fact that every
LGU is an LPU).
If $\psi_0$ is some simpler state, such as a product state or the ground state of an exactly solvable Hamiltonian, then it may be easier to evaluate the expectation value
of $\Bigl( U^\dagger O U \Bigr) $ in $\psi_0$.  This approach is used in Ref.~\onlinecite{bhv}, for example, to study the generation of correlations and topological order.

Some formal steps need to be taken for an infinite system to define the limits; one must use the locality of the definition for a finite system to show the existence of certain limits.  Since our definition is simply the evolution under a time-dependent Hamiltonian obeying a Lieb-Robinson bound, the results in Ref.~\onlinecite{thermolimit} allow one to work directly in the thermodynamic limit using the C$^*$-algebraic definition of the quasi-local algebra.  Alternately, one can work throughout with finite systems and use the fact that the Lieb-Robinson bounds are uniform in the system size.

\subsection{Locality-Preserving Unitaries}
We define an LPU as follows.  The definition is motivated by the analogy: an LPU has the same relation to an LGU as a QCA does to a quantum circuit. That is, just as in a QCA we kept one property of the quantum circuit (that it increased the diameter of the support of operators by a bounded amount) and removed the others, for an LGU we will keep the property Eq.~(\ref{LRbound}) and remove the others.  Thus, for a finite system, an LPU with control function $K$ will be defined to be a unitary $U$ such that for any set $A$ and for any operator $O_A$  supported on $A$, and for any $R$,
there is an operator $O_{b_R(A)}$ supported on $b_R(A)$ such that
\be
\label{LRboundLPU}
\Vert U O_A U^\dagger - O_{b_R(A)} \Vert  \leq \Vert O_A \Vert \sum_{i \in A} K(R+{\rm dist}(i,\overline A)),
\ee
for some ``control function" $K(R)$ such that $\lim_{R \rightarrow \infty} K(R)=0$ and also such that there is some other operator $O'_{b_{R(A)}}$ such that
\be
\label{LRboundLPUdagger}
\Vert U^\dagger O_A U - O_{b_R(A)} \Vert  \leq  \Vert O_A \Vert  \sum_{i \in A} K(R+{\rm dist}(i,\overline A)).
\ee
In this case, we absorb the constant $2C$ that appeared in Eq.(~\ref{LRbound})  into the definition of $K(R)$.
An analogous definition can also be made for infinite systems in terms of automorphisms, where we replace $U O_A U^\dagger$ by an automorphism $\alpha_{LPU}(O_A)$, or by considering families of such unitaries $U$ on finite systems of increasing size.

We can consider various choices of the control function $K(R)$.  For example, we can consider the LPU such that $K(R)=0$ for sufficiently large $R$.  The set of all such LPU is equivalent to the set of QCA and this set forms a group; for example, the composition of a QCA with range $R_1$ with another QCA with range $R_2$ is a QCA with range $R_1+R_2$.
For $d$-dimensional lattice systems,
the set of LPU such that $K(R)$ decays exponentially also forms a group under composition, as does the set such that $K(R)$ decays superpolynomially.
To see this, consider first the exponential decay case.  Consider the composition of two automorphisms $\alpha_2(\alpha_1(O_A))$.
Consider a given set $A$, given $O_A$, and given $R$.  Approximate $\alpha_1(O_A)$ by an operator $O_{b_{R/2}(A)}$ supported on the set of sites within distance $R/2$ of $A$, up to error $\Vert O_A \Vert \sum_{i \in A} K(R/2+{\rm dist}(i,\overline A))$.  
Then, approximate $\alpha_2(O_{b_{R/2}(A)})$ by an operator supported on the set of sites within distance $R$ of $A$.  Since this set is within distance $R/2$ of the set $b_{R/2}(A)$, the error is bounded by
\be
\label{err1}
\Vert O_{b_{R/2}}(A) \Vert \sum_{i \in b_{R/2}(A)} K(R/2+{\rm dist}(i,\overline b_{R/2}(A))).
\ee
We estimate $\sum_{i \in b_{R/2}(A)} K(R/2+{\rm dist}(i,\overline b_{R/2}(A)))$.
The sum over $i$ in $b_{R/2}(A)$ can be broken into two sums: a sum over $i \in A$ and a sum over $i \not \in A$.  The first sum is bounded by
$\sum_{i \in A} K(R+{\rm dist}(i,\overline A)$.  As for the second sum, let $N_0$ be the number of sites on the boundary of $A$.  
Let $M_0$ denote the number of sites on the boundary of $b_{R/2}(A)$, let $M_1$ denote the number of sites in $b_{R/2}$ which are a distance $1$ from the boundary and so on.  Then, $M_0 \leq  N_0 O(R/2+1)^d$, and $M_k \leq N_0 O(R/2+1-k)^d$, for $0 \leq k \leq R/2)$.  So,
\be
\label{err2}
\sum_{i \in b_{R/2}(A)} K(R/2+{\rm dist}(i,\overline b_{R/2}(A))) \leq \sum_{i \in A} K(R/2+{\rm dist}(i,\overline A)) + N_0 \sum_{k=0}^{R/2} O(R/2+1-k)^d K(R/2+k).
\ee
For $K(R)$ either bounded by an exponentially decaying function of $R$ or by a superpolynomially decaying function of $R$, the sum of Eqs.~(\ref{err1},\ref{err2}) is bounded by an exponentially or superpolynomially decaying function.
This shows the claimed group property.

\section{Topological Order and Intrinsic Topological Order}
\label{nointo}
As mentioned in the introduction, we often consider two gapped, local quantum Hamiltonians $H_0,H_1$ to be equivalent if there is a continuous path of gapped, local Hamiltonians connecting them.  As discussed above, this implies that, by quasi-adiabatic continuation, we can map the ground state of one into the other using an LGU.  Conversely, if there is an LGU $U_t$ that maps the ground state of $H_0$ into $H_1$, then there is a continuous path of gapped local Hamiltonians connecting them: consider the path $U_s H_0 U_s^\dagger$ for $0 \leq s \leq t$; then linearly interpolate between $U_t H_0 U_t^\dagger$ and $H_1$.  So, one often defines that a state is ``trivial" if it can be mapped to a product state by an LGU.

So, we propose the following definition:
\begin{definition}
\label{ITOdef}
A state is ``nontrivial and has intrinsic topological order" if it cannot be mapped to a product state by an LPU while a state is ``nontrivial without intrinsic topological order" if it can be mapped to a product state by an LPU but not an LGU.
\end{definition}

Similarly,
\begin{definition}
\label{ITOHdef}
A Hamiltonian $H$ which is a sum of commuting terms is ``nontrivial and has intrinsic topological order" if it cannot be mapped to a Hamiltonian $H_{triv}$ which is a sum of commuting terms with disjoint support by an
LPU while a Hamiltonian is ``nontrivial without intrinsic topological order" if it can be mapped to such a Hamiltonian by an LPU but not an LGU.
\end{definition}
We allow the Hamiltonian $H_{triv}$ to be a sum of terms with disjoint support.  So, that allows not just Hamiltonians which are a sum of terms supported on different sites but also Hamiltonians which have dimer ground states such as a Hamiltonian for a spin-$1/2$ system that is a sum of projectors on spins $2x,2x+1$, projecting onto the triplet states.
These definitions suffice for infinite systems, while for finite systems we need to consider families of Hamiltonians on systems of increasing size and impose uniform bounds on the LPU or LGU to obtain an appropriate definition.

These definitions can naturally be generalized to the case of symmetries, as we can impose the same symmetry on the LPU or LGU.

Interestingly, the proofs that the toric code is nontrivial in that no LGU maps its ground state to a product state (see Ref.~\onlinecite{bhv} for the code on a torus and Ref.~\onlinecite{leshouches} for other topologies where the proof does not use the ground state degeneracy) only use one property of an LGU: that an LGU is an LPU, though that terminology was not used there.  Hence those proofs immediately extend to proofs that the toric code has intrinsic topological order according to this definition.  Conversely, later when we discuss the application of QCAs to classifying phases with symmetry, we will see some phases which are nontrivial without intrinsic topological order.  So, our definition may be a way to propose a precise definition which accords with the informal usage in the literature.

To illustrate one important difference between phases with and without intrinsic topological order, consider the Hamiltonian for a spin-$1/2$ system which is $H=\sum_i S^z_i$.  This is a gapped Hamiltonian, whose ground state has all spins pointing down.  We can create a single excitation by acting on the ground state with a local operator, namely $S^x_i$ or $S^y_i$.  Suppose we define a new Hamiltonian by conjugating this Hamiltonian by some LPU $U_{LPU}$.  The resulting Hamiltonian will still local, and one can still create a single excitation with a local operator, namely the operator that is obtained by conjugating $S^x_i$.  Conversely, in the case of the toric code on an infinite system, while the Hamiltonian is local, there is no local operator that creates a single excitation: one needs a string going off to infinity.
 The reader may wonder whether there exists an LPU $U_{LPU}$ that is not also an LGU, and also whether
the existence of such an LPU would imply that $U_{LPU} H L_{LPU}^\dagger$ is nontrivial; these issues are discussed later.

While we present examples showing that this definition works reasonably for symmetry protected phases, it seems that the definition is not appropriate for integer quantum Hall phases.  We would like to say that those are phases without intrinsic topological order, but it seems that they cannot be transformed to an $H_{triv}$ by an LPU.

\section{Complications With Intrinsic Topological Order}
\label{complications}
A key step in the calculation in the previous section was that we could add a term to the Hamiltonian supported near the puncture to ``heal" the puncture and restore the property that the Hamiltonian has a spectral gap.
In the rest of the paper, we generalize the torus trick beyond the case of free fermions.  However, if we over-generalize, then we find that healing the puncture in this way will not always be possible.  This is related to a difference between homotopy invariants and locally computable invariants in this case.

\begin{figure}
\includegraphics[width=3in]{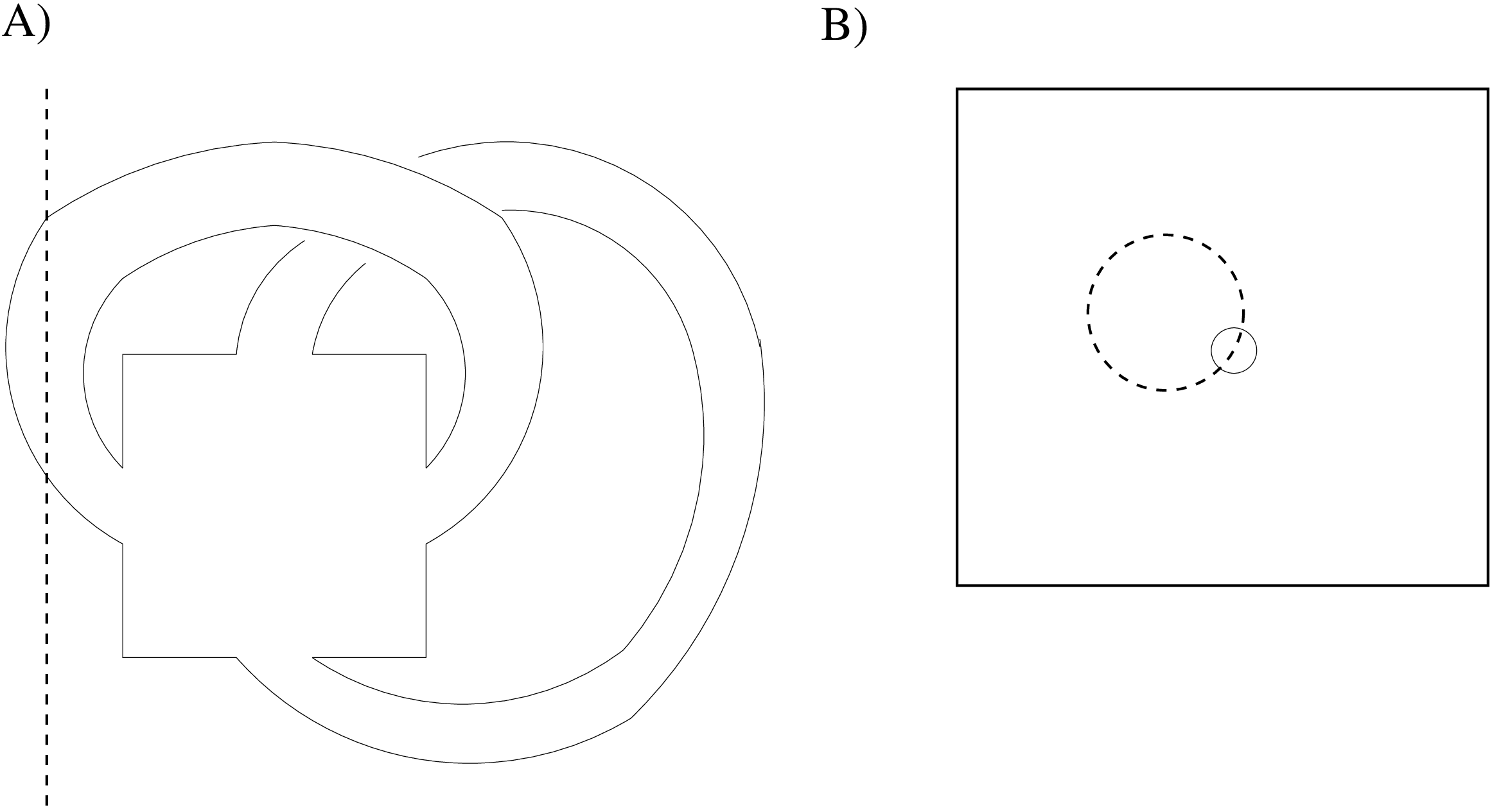}
\caption{A) Immersed punctured torus.  Vertical dashed line shows boundary between two copies of toric code (on right) and vacuum (on left).  B) Pullback to punctured torus.  Dashed circle is inverse image of dashed line, thin solid circle is puncture.}
\label{FigFolded}
\end{figure}

Consider the system shown in Fig.~\ref{FigFolded}A.  This is a two-dimensional system.  
On the right-half of the system, we have two copies of the toric code\cite{tc}.  On the left half of the system, we have a trivial system.  The vertical dashed line indicates the separation between the two halves of the system.
The solid line denotes an immersed punctured torus.

Such a system can be defined by a local Hamiltonian without having gapless edge modes at the vertical line.  To see this, consider a uniform system with one copy of the toric everywhere.  Then ``fold" the system over at the vertical line: place $(x,y)$ coordinates on the system, taking the vertical line at $x=0$, and relabel the coordinates of the site so that sites with $x>0$ are left unchanged but those with $x<0$ are mapped by $x \rightarrow -x$.
Such a procedure is not specific to the toric code; we could do this for any gapped Hamiltonian.  However, if the original Hamiltonian is not invariant under spatial reflection, then the result  is not two copies of the same system on the right-hand side, but rather two different systems related by spatial reflection.

So, this example shows that locally computable invariants cannot distinguish between two copies of the toric code and the vacuum.  However, two copies of the toric code is not homotopy equivalent to the vacuum, as can be seen by for example the argument in \cite{bhv} for a finite system on a torus or the argument of \cite{leshouches} more generally.

In Fig.~\ref{FigFolded}B we show the Hamiltonian on the punctured torus. Identifying opposite sides of the square gives a torus.  The dashed line is the inverse image of the vertical line in Fig.~\ref{FigFolded}A, while the thin solid circle is the puncture.  Suppose we add terms to the Hamiltonian supported near the puncture to  ``heal" the puncture.  We will still be left with vacuum inside the solid line and two copies of the toric code outside the solid line.  The resulting Hamiltonian does not have a gap.
One may say that even though we healed the puncture at the solid line, there still is a hole inside the dashed line that we have not yet healed.

In this particular case, we can create a gap in the Hamiltonian by adding terms supported near the dashed line.  This leads to only a slight further violation of locality (we can use the same trick as before to shrink the solid line to a point, slightly stretching the distance between other points).  So, in this case we can gap the Hamiltonian while maintaining locality.
Similarly, if we had positioned the immersed punctured torus in Fig.~\ref{FigFolded}A slightly further to the right, there would be no problem.  However, if we instead slide the immersed torus further to the left, eventually there will be a problem: the solid line will become sufficiently large that it cannot be shrunk to a point without severely violating locality and we will instead have to discontinuously jump to a different method of gapping the system out.
Further, unlike in the free fermion case where the discontinuous change in how we heal the puncture was confined to a small region and did not change the K-theory class, here the change does change the homotopy class.

We leave it as an open problem whether for every Hamiltonian which is a sum of commuting local projectors obeying the conditions called TQO-1 and TQO-2 in Ref.~\onlinecite{bhv} and for every finite region $Z$ there exists a periodic Hamiltonian which is a sum of commuting local projectors and which obeys TQO-1 and TQO-2 and which agrees with the original Hamiltonian on $Z$.

\section{The Torus Trick for Quantum Cellular Automata}
We now describe the application of the torus trick to QCA.  We consider an aperiodic QCA $\alpha$ with some given range $R$.  This QCA could be finite or infinite.
We use the same immersions $f_{z,L}$ as before.
We define the same set of sites as before on the punctured torus, by taking the pre-image of the set of sites in the image of the immersion.
Each site in the pre-image will have a Hilbert space with the same dimension as the Hilbert space on the corresponding site in the image.

Later, we will consider QCA obeying certain symmetry constraints.  As in the case of free fermions, the torus trick can be applied to a QCA with symmetry and the result is a periodic QCA with symmetry.  That is, again the symmetry ``goes along for the ride".

Let $X$ be the set of sites in the pre-image such that the corresponding site in the image is more than distance $R$ away from the image of the puncture.
To pullback $\alpha$,
we will define a map $\alpha_{pt}$ that is a homomorphism from the algebra of operators supported on sites in $X$ to the algebra of all operators on the punctured torus.  Note that $\alpha_{pt}$ need not be an automorphism; constructing an automorphism will be the next step to ``heal the puncture".
We take $L$ sufficiently large that
\be
\label{LsuffL}
c_{inj} C L >2R,
\ee
where $C$ is the constant appearing in Eq.~(\ref{stretch}).
Consider any site $\tilde i \in X$
and any operator $O_{\tilde i}$ supported on that site.
The immersion is injective within a distance $c_{inj} L$ of $\tilde i$, and so by Eq.~(\ref{LsuffL}) it is injective within a distance greater than $2R$ of the image of $\tilde i$.
So, we define $\alpha_{pt}(O_{\tilde i})$ in the natural way by pulling back $\alpha(O_{f_{z,L}}(\tilde i))$.  To state it explicitly, define an isomorphism $\beta_{\tilde i}$ from the algebra of operators on the Hilbert space on sites within distance $c_{inj} L$ of $\tilde i$ to the algebra of operators on the image of those sites in the natural way: the immersion is one-to-one so we map an operator supported on a site to the corresponding operator on the image of that site.  Then, we define $\alpha_{pt}(O_{\tilde i})=\beta^{-1}_{\tilde i} \circ \alpha \circ \beta_{\tilde i}(O_{\tilde i})$.

For $\tilde i \in X$, one can verify that $\alpha_{pt}$ defines a homomorphism of the algebra ${\cal A}_{\tilde i}$ of operators on $\tilde i$ to the algebra of all operators.
We define $\alpha_{pt}(O)$ for any operator $O$ which is a product of operators $O_{\tilde i}$ for $\tilde i \in X$ by:
\be
\alpha_{pt}(O_{\tilde i_1} O_{\tilde i_2} ...)=\alpha_{pt}(O_{\tilde i_1}) \alpha_{pt}(O_{\tilde i_2}) ...
\ee
and we extend this to all operators supported on $X$ by linearity.
To show that this is well-defined independently of the order of sites, and to show that $\alpha_{pt}$ is a homomorphism, we need to verify that for $\tilde i \neq \tilde j$ we have that
$[\alpha_{pt}(O_{\tilde i}),\alpha_{pt}(O_{\tilde j})]=0$.  For ${\rm dist}(\tilde i,\tilde j)>2R/C$, this follows immediately because the supports 
of $\alpha_{pt}(O_{\tilde i})$ and $\alpha_{pt}(O_{\tilde j}$ are disjoint.  For ${\rm dist}(\tilde i,\tilde j)\leq 2 R/C$, by assumption we have that the immersion is injective on a set that contains the supports $\alpha_{pt}(O_{\tilde i})$ and $\alpha_{pt}(O_{\tilde j})$ so  the commutator can be computed by pushing forward $O_{\tilde i}$ and $O_{\tilde j}$ to $O_i$ and $O_j$, then applying $\alpha$ and taking the commutator, and pulling back; however, since
$[\alpha(O_i),\alpha(O_j)]=0$, the commutator is zero.

Having defined the homomorphism $\alpha_{pt}$, we extend this homomorphism to an automorphism $\alpha'_{torus}(\alpha,Z)$ of the algebra of operators in any arbitrary way.  This extension ``heals the puncture".  To see that such an automorphism exists, note that $\alpha_{pt}$ is a homomorphism of an algebra ${\cal A}_X$ of operators on $X$ which has no central elements.  So, $\alpha_{pt}({\cal A}_X)$ is isomorphic to a matrix algebra, so it defines some tensor product decomposition of the algebra of operators.

Note that for operators $O$ in $X$, we have that $\alpha'_{torus}(O)$ has support on the set of sites within distance $R/C$ of the support of $O$.  Since the set of sites not in $X$ has a diameter bounded by a constant factor times $R$, we see that $\alpha'_{torus}(\alpha,Z)$ has a range that is only a constant factor larger than $R$.

For $\Ls$ sufficiently large compared to $R$ ,we can then ``unfurl" $\alpha'_{torus}(\alpha,Z)$ to a periodic system $\alpha'(\alpha,Z)$ on $R^d$, defining this unfurling in the natural way: for every site $i$ in $R^d$, we define $\alpha'(\alpha,Z)$ acting on $O_i$ by pushing forward to the torus, applying $\alpha'_{torus}$, and pulling back, using the fact that the covering map from the plane to the torus is one-to-one on sets of diameter small enough compared to $\Ls$ to define the pullback.
Conversely, given a periodic QCA, we can furl it back to a QCA on the torus assuming that the range $R$ is sufficiently small; $R < \Ls/4$ suffices and in general we will assume $R< \Ls/4$ for any QCA on a torus.

A natural question is whether a similar trick can be developed for LPU with exponentially decaying or superpolynomially decaying control function.  This may be a subtle issue, as is understanding whether such an LPU can be approximated by a QCA with finite range $R$.

The torus trick lets us prove the following theorem analogous to {\bf 2} in theorem \ref{mainFFlemma}.  In order to prove a result analogous to {\bf 1}, it may be necessary to have a better understanding of LPU with  with exponentially decaying or superpolynomially decaying control function in order to smoothly vary the
LPU over space in a way analogous to how the free fermion Hamiltonian was varied over space.
First we define
\begin{definition}
Consider two QCAs $\alpha_0$ and $\alpha_1$ defined on a torus with $R<\Ls/4$.  We say that they are stably homotopy equivalent if we can find tensor in additional degrees of freedom to both $\alpha_0$ and $\alpha_1$ so that $\alpha_0 \otimes I$ can be deformed to $\alpha_1 \otimes I$ by a continuous path of QCA with range smaller than $\Ls/4$.
\end{definition}
Note that as in the case of free fermions, even if we do not include moving the location of sites in the ambient space as part of the definition, we can obtain this by tensoring in additional degrees of freedom.  

\begin{definition}
We say that a QCA $\alpha$ agrees with a QCA $\alpha_0$ on $Z_0$ if the action of $\alpha$ on any operator supported on $Z_0$ is the same as that of $\alpha_0$ on that operator.  We say that a QCA $\alpha$  interpolates between $\alpha_0$ on $Z_0$ and $\alpha_1$ on $Z_1$ if it agrees with $\alpha_0$ on $Z_0$ and agrees with
$\alpha_1$ on $Z_1$.
\end{definition}

\begin{theorem}
\label{mainQCATlemma}
Consider any two QCAs $H_0,H_1$ with range bounded by $R$, and consider any two hypercubes $Z_0,Z_1$ with linear size $L$ with $L$ sufficiently large compared to $R$.
 If $\alpha'_{torus}(\alpha_0,Z_0)$ is not stably homotopy equivalent to $\alpha'_{torus}(\alpha_1,Z_0)$, then there is
no QCA $\alpha$ with given range $R$ which interpolates between $\alpha_0$ on $Z_0$ and $\alpha_1$ on $Z_1$.
\begin{proof}
 Suppose such an interpolating QCA does exist.  Then, consider a path of hypercubes $Z_s$ that starts at $Z_0$ and ends at $Z_1$ by continuously sliding and translating the hypercube.  Such a path of hypercubes defines a path of $\alpha'_{torus}(H,Z_s)$.  This path  may have discontinuities. However, for sufficiently large $L$ compared to $R$, the QCA is stably equivalent across these discontinuities, as these discontinuities all happen near the puncture; that is, the action of the QCA on observables sufficiently far from the puncture is smooth across the discontinuities.  So, across a discontinuity, we stabilize by adding additional sites near the puncture so that the dimensions of the QCAs match.  Then, we can pick any choose of action of the QCA on the observables near the puncture without violating the bound on the range, and so the problem near the puncture is that of classifying zero dimensional unitaries and since the dimensions of the unitaries match there are no obstructions to finding a path between any two unitaries. 
\end{proof}
\end{theorem}

\section{The Torus Trick For Systems Without Intrinsic Topological Order}
We are primarily interested in QCA in their role for classifying systems without intrinsic topological order following definition \ref{ITOHdef}.  However, the classification of Hamiltonians without intrinsic topological order is not the same as that of QCA.  Given some Hamiltonian $H_{triv}$, we can define a group $G_{sym}$ of LPUs which preserve that Hamiltonian.  Then, the classification of Hamiltonians which can be mapped to $H_{triv}$ by an LPU is the classification of $G_{LPU}/G_{sym}$, where $G_{LPU}$ is the group of LPU.

However, a torus trick can be directly developed for Hamiltonians without intrinsic topological order.  If the Hamiltonian is obtained by acting on $H_{triv}$ with a QCA (rather than just an LPU) then the terms have bounded range and we can pullback the Hamiltonian to the torus in the natural way.  Unlike the case with intrinsic topological order, it is possible to add terms near the puncture to heal the puncture and obtain a gapped periodic system, as follows from the ability to heal the puncture for QCAs.
This allows us to prove a result analogous to theorem \ref{mainQCATlemma} for such systems without intrinsic topological order.  This will be discussed elsewhere, as will be the classification of periodic systems without intrinsic topological order.

\section{Classification of Quantum Cellular Automata}
We now present various results on the classification of QCA.  The first subsection reviews previous results of Ref.~\onlinecite{QCA} on one-dimensional QCA.  The results there are applicable to either the periodic or aperiodic case, so there is no need for the torus trick.  Next, we consider the case of one dimension without symmetry; again the torus trick is not required.  The subsequent subsections consider classification of periodic QCA  in higher dimensions; only partial results are presented here.

\subsection{In One Dimension}
\label{inoned}
We review briefly Ref.~\onlinecite{QCA}.  The result will be that QCAs are classified by a positive rational, which fully classifies both the homotopy invariants and the local invariants.  Before giving the result, we give some intuitive idea by an example.  Consider a system with a $p$ dimensional Hilbert space on each site.  Consider the QCA $\alpha_L$ which ``shifts" any operator one site to the left, mapping any operator $O_i$ supported on site $i$ to an operator supported on site $i-1$ under some automorphism of the algebra of $p$-by-$p$ matrices.  Consider another system with a $q$ dimensional Hilbert space on each site, and a QCA $\alpha_R$ which shifts operators in this system one site to the right.  Then $\alpha_R \otimes \alpha_L$ will have index
$p/q$.

Also, before reviewing the result, we make some comments regarding the application of this invariant to higher dimensional QCAs constructed by the torus trick as defined here, remarking that the torus trick ``forgets" this invariant for $d>1$.  Consider a QCA on a $d$-dimensional torus, for $d>1$.  We can choose to regard this as a QCA on a one-dimensional torus by ignoring the locality properties in $d-1$ of the coordinates and then we can define a one-dimensional invariant.  Such a QCA certainly exists and certainly the one-dimensional invariant represents an invariant of such QCAs.  However, a QCA with a nontrivial value of this invariant cannot arise via the torus trick from a system with $d>1$, because each $T^{d-1}\times [0,1]$ in the immersion gives a decomposition of $R^d$ into three regions:
 the interior, $T^{d-1} \times [0,1]$, and the exterior.  Having a nontrivial value of the one-dimensional invariant would imply
 a nontrivial value of the invariant for a three site system with the three sites consisting of these three regions and this is impossible because two of the regions are finite dimensional.

A key concept in defining the index is the idea called a ``support algebra" in that paper and originally called an ``interaction algebra" in Ref.~\onlinecite{intalg}.  Consider two sets of sites $Z_1$ and $Z_2$ with ${\cal A}_1$ and ${\cal A}_2$ being the algebras of operators on those two sites respectively.  Let ${\cal A}$ be a subalgebra of ${\cal A}_1 \otimes {\cal A}_2$.  Consider an orthonormal basis of operators $e_i$ for ${\cal A}_2$.  Then, every operator in ${\cal A}$ can be decomposed as $\sum_i a_i \otimes e_i$ with $a_i \in {\cal A}_1$.
The ``interaction algebra of ${\cal A}$ on $Z_1$" (equivalently, the ``interaction algebra of ${\cal A}$ on ${\cal A}_1$") is defined to be the subalgebra of operators on ${\cal A}_1$ generated by all the $a_i$ arising from such a decomposition.

Consider a QCA $\alpha$.
By coarse-graining if needed, we can assume that the QCA has range $R=1$.  Let sites be labeled by integers $i$ with corresponding algebras ${\cal A}_i$ of operators on site $i$.  For integer $x$, define ${\cal R}_{2x}$ to be the interaction algebra of $\alpha({\cal A}_{2x} \otimes {\cal A}_{2x+1})$ on ${\cal A}_{2x-1} \otimes {\cal A}_{2x}$ and let 
${\cal R}_{2x+1}$ to be the interaction algebra of $\alpha({\cal A}_{2x} \otimes {\cal A}_{2x+1})$ on ${\cal A}_{2x+1} \otimes {\cal A}_{2x+2}$.
A crucial fact that one can check is that the algebras ${\cal R}_i$ commute with each other.  This follows trivially for ${\cal R}_i, {\cal R}_j$ for $|i-j|>1$.  For $i=2x+1$ and $j=2x+2$, this follows from the fact that $\alpha({\cal A}_{2x} \otimes {\cal A}_{2x+1})$ commutes with 
$\alpha({\cal A}_{2x+2} \otimes {\cal A}_{2x+3})$ and from the fact that
$\alpha({\cal A}_{2x} \otimes {\cal A}_{2x+1})$ is supported on sites $2x-1,2x,2x+1,2x+2$ and $\alpha({\cal A}_{2x+2} \otimes {\cal A}_{2x+3})$ is 
supported on sites $2x+1,2x+2,2x+3,2x+4$ and the intersection of these supports is sites $2x+1,2x+2$.

Let $d(i)$ be the dimension of the Hilbert space on site $i$.  It is shown in Ref.~\onlinecite{QCA} that the ${\cal R}_i$ have no central elements and hence are isomorphic to matrx algebras of some dimension $r(i)$.  Further, it is shown that
\be
\label{recall}
d(2x) d(2x+1)=r(2x)r(2x+1)
\ee
and
\be
\label{recall2}
r(2x+1) r(2x+2)=d(2x+1) d(2x+2).
\ee
Hence,
\be
\frac{r(2x)}{d(2x)}=\frac{r(2x+1)}{d(2x+1)}=\frac{r(2x+2)}{d(2x+2)}.
\ee
This ratio is a positive rational number, which is the index of $\alpha$.  The above equation shows that it is a local invariant as it is the same for all $x$.

\subsection{Invariants In One Dimension With Symmetry}
Suppose the unitary $U_{QCA}$ commutes with some global symmetry.  This could be a discrete symmetry group $G$, meaning that for each site $i$ there is a homomorphism from group elements $g \in G$ to unitaries $g_i$ supported on site $i$, such that for any $g$, the product $\prod_i g_i$ commutes with $U_{QCA}$.  In the case of a continuous symmetry, we can consider a homomorphism from the Lie algebra to operators $q_i$ supported on site $i$ such that $\sum_i q_i$ commutes with $U_{QCA}$.
The following result for systems with symmetry will be useful in several problems, so we state it as a lemma here.
\begin{lemma}
Consider a QCA $\alpha$ on an infinite one-dimensional system.  Assume that the range $R=1$.  Define ${\cal A}_i$ and ${\cal R}_i$ as before.
For given $g\in G$, let $\alpha_g$ be the QCA that conjugates the algebra on each site by $g_i$ for unitaries $g_i$ supported on site $i$.  Assume that $\alpha \circ \alpha_g = \alpha_g \circ \alpha$.  Then, for any $x$,
$g_{2x-1} g_{2x}$ conjugates the algebra ${\cal R}_{2x}$ to itself and also conjugates ${\cal R}_{2x-1}$ to itself.

Further, we can decompose for any $x$
\be
\label{factor}
g_{2x} g_{2x-1}=g^r_{2x} g^r_{2x-1}
\ee
for some unitaries $g^r_i \in {\cal R}_{i}$.
\begin{proof}
Recall that ${\cal R}_{2x}$ is the interaction algebra of $\alpha({\cal A}_{2x} \otimes {\cal A}_{2x+1})$ on ${\cal A}_{2x-1} \otimes {\cal A}_{2x}$ and 
${\cal R}_{2x+1}$ is the interaction algebra of $\alpha({\cal A}_{2x} \otimes {\cal A}_{2x+1})$ on ${\cal A}_{2x+1} \otimes {\cal A}_{2x+2}$.
Consider an operator $O$ in ${\cal A}_{2x} \otimes {\cal A}_{2x+1}$.  Note that
$\alpha_g \circ \alpha(O)=\alpha \circ \alpha_g(O)$.  Since $\alpha_g(O)$ is in ${\cal A}_{2x} \otimes {\cal A}_{2x+1}$, it follows that
$\alpha \circ \alpha_g(O)$ can be decomposed as a sum of products of operators of operators in ${\cal A}_{2x-1} \otimes {\cal A}_{2x}$
with operators in ${\cal A}_{2x+1} \otimes {\cal A}_{2x+2}$.  So, $\alpha_g \circ \alpha(O)$ can be decomposed in the same way, so 
$g_{2x-1} g_{2x}$ conjugates the algebra ${\cal R}_{2x}$ to itself as claimed.  The claim that $g_{2x-1} g_{2x}$ conjugates ${\cal R}_{2x-1}$ to itself is proven similarly.

Then since $g_{2x-1} g_{2x}$ conjugates $R_{2x}$ to itself and conjugates $R_{2x-1}$ to itself, and ${\cal R}_{2x-1} \otimes {\cal R}_{2x}={\cal A}_{2x-1} \otimes {\cal A}_{2x}$, Eq.~(\ref{factor}) follows.
\end{proof}
\end{lemma}

Given this lemma, the map from $G$ to $g^r_i$ gives some {\it projective} representation of the the group $G$.  This representation is projective, because we can absorb a phase factor into $g^r_x$ and the opposite phase factor into $g^r_{2x-1}$ so that the phase of $g^r_i$ can be chosen arbitrarily.
This representation need not be irreducible.  Let us consider some specific examples.  Consider a system of spin-$1/2$ particles, and consider a QCA $\alpha$ that shifts by distance $R$ to the right.  If $R=0$, then one finds that the $g^r_i$ are a spin-$1/2$ representation.  If $R=1$, then ${\cal R}_{2x}$ is trivial and so the representation $g^r_{2x}$ is a trivial (spin-$0$) representation, while the representation $g^r_{2x-1}$ is a sum of a spin-$0$ and a spin-$1$ representation.  In general for odd $R$, after coarse-graining so that we can apply the above lemma, we find that the $g^r_{2x}$ give an integer spin-representation, while it is not an integer spin representation for even $R$.

How can this representation change under deformation of the QCA?  Let us allow the range to increase under this deformation (currently we are not working on a torus but rather on a line).  Take some given QCA $\alpha_0$ with range $R_0$ and follow some path of QCA $\alpha_s$, with range $R_s$, with $R_s \leq R$ for all $s$.  Then, we can block the system into blocks of size $R$ and apply the above result.  Then, the representation does not change along the path.

However, under this coarse-graining, or equivalently under tensoring in additional degrees of freedom, the representation may change.
That is, if for a given QCA $\alpha$, the map from $g$ to unitaries $g^r_{2x}$ defines a given projective representation $r$, and if the identity QCA $I$ has a representation $r'$, then this representation $r$ remains constant along the path, but if we tensor in additional degrees of freedom representation $R$, then now $\alpha \otimes I$ has representation $r \otimes R$, while $I \otimes I$ has representation $r' \otimes R$.  So, to decide if two QCA are different, we must determine if $r \otimes R$ is distinct from $r' \otimes R$ for all $R$.  Of course, if the dimension of $r$ differs from that of $r'$, this describes a different QCA and the ratio of the dimensions ${\rm dim}(r)/{\rm dim}(r')$ is the index of Ref.~\onlinecite{QCA}.  If the dimensions agree,  for a finite group, then if $r$ is an ordinary representation $r \otimes R \cong r' \otimes R$ for $R$ being the regular representation\cite{moverflow} (note that the identity QCA has an ordinary representation by definition).  One can show that a one-dimensional QCA with finite symmetry group is trivial if and only if they have the index (as in Ref.~\onlinecite{QCA}) equals $1$ and if the representation $r$ is an ordinary representation (more precisely, if the projective representation determined by the unitaries $g^r_i$ can be lifted to an ordinary representation).

In the case of a compact, connected Lie group\cite{moverflow}, for any choice of a finite-dimensional $R$ we have that $r\otimes R \cong r' \otimes R$ only if $r \cong r'$.  Now, a QCA is trivial if and only if the index equals $1$ and the representation given by the map from $G$ to unitaries $g^r_{2x}$ is isomorphic to the representation given by the map from $G$ to unitaries $g_{2x}$.

\subsection{QCAs and Symmetry Protected Phases}
Let us consider a Hamiltonian $H_{triv}$ for a spin-$1/2$ system which is a sum of projectors on spins $2x,2x+1$, projecting onto the triplet states.  Act on this with a QCA that shifts by an odd number of sites.  This defines a new Hamiltonian.  The original Hamiltonian is invariant under a group $G_{sym}$ of LPUs which preserve that Hamiltonian.  This group contains only QCAs with even shift.  So, the QCAs with odd shift produce nontrivial phases.  For finite groups, we obtain a classification by the projective representations (i.e., the second cohomology) as emphasized by many other authors.

\subsection{In Two Dimensions With Discrete Symmetry}
We now consider the case of classifying QCA in two dimensions with symmetry.  We emphasize that this may be only a partial classification.  There may exist, for example, nontrivial QCA even without any symmetry requirement.
Our argument closely parallels that of Ref.~\onlinecite{sptU}.  We will emphasize a few differences from their argument.

We only sketch the argument.  First, if we start with a finite aperiodic system, use the torus trick to consider infinite periodic QCA in two dimensions.  Put $x$ and $y$ coordinates on the plane, 
and define the unit cell so that it is parallel to these axes.
Pick a group element $g$ and integer $X$ and define a QCA $g_X$ to conjugate all observables  by the action of group element $g$ on all sites with $x$-coordinate $-\infty <x \leq X$.
Consider the QCA $\alpha^{-1} \circ g_0 \circ \alpha$.  For sites sufficiently far to the right of the vertical line $x=0$, this QCA acts as the identity, while for sites sufficiently far to the left of the line, this QCA conjugates all observables by the action of group element $g$ (i.e., for sites sufficiently far to the left, the action of this QCA is the same as that of $g_0$).  So, $g_{-R}^{-1} \circ \alpha^{-1} \circ g_0 \circ \alpha$ acts only near the vertical line (i.e., it is a one-dimensional QCA).
Call this QCA $\beta_g$.  Note that the $\beta_g$ form a representation of the group: $\beta_g \circ \beta_h = \beta_{gh}$.

A similar one-dimensional QCA is constructed in Ref.~\onlinecite{sptU}, but we use a slightly different construction that does not require referring to low energy or boundary degrees of freedom but is directly constructed from the QCA.
Then, as in Ref.~\onlinecite{sptU}, we exactly write such a one-dimensional QCA as a matrix product operator\cite{mpo} with a bounded bond dimension $k$; the bound on the bond dimension follows from the bound on the range $R$. 
We now review this argument.
Consider the product $\beta_g \circ \beta_h$.  The product of the corresponding matrix product operators has, when written naively, a bond dimension $k^2$.  However, because it is equal to $\beta_{gh}$ up to a phase it is possible to reduce the bond dimension to $k$ by left-multiplying the matrices by some matrix $V_{g,h}$ and right-multiplying by some $V_{g,h}^\dagger$, where $V_{g,h}$ is an isometry from a $k$-dimensional space to the $k^2$-dimensional space; note that $V_{g,h}^\dagger V_{g,h}$ is a projection.
Then, this left- and right-multiplication by $V_{g,h}$ gives the matrices of $\beta_{gh}$ up to a phase.
We put the subscripts $g,h$ on $V$ to denote that each choice of $g,h$ gives some corresponding matrix $g$.
There is a phase ambiguity in this isometry; generically it is possible to pick a canonical form\cite{canon} for the matrix product operator in which this is the only freedom, though in non-generically the symmetry may be enhanced.  Now, consider the product of three matrix product operators, $\beta_f, \beta_g, \beta_h$.  We have that
$V_{fg,h} (V_{f,g} \otimes I)$ is equal to $V_{f,gh} (I \otimes V_{g,h})$ up to a phase $\exp(i\phi(f,g,h))$.
  This phase is an element of the third cohomology $H^3(G,U(1))$;
note that if we re-define the $V_{f,g}$ by multiplying by a phase $\exp(i \theta_{f,g})$, this changes the phase $\exp(i \phi(f,g,h))$ by a coboundary.
One can also consider different ways to combine the product of four matrix product operators to verify that the phase is a cocycle.
The QCA $\alpha={\rm identity}$ gives a trivial element of cohomology.

\subsection{In Arbitrary Dimension With Continuous $U(1)$ Symmetry}
We can define another invariant in the case that there is a continuous global $U(1)$ symmetry.  We define this invariant for a QCA on a $d$-dimensional torus, as can be constructed using the torus trick.  Suppose that for every site $i$ there is an operator $q_i$ which has integer eigenvalues such that
$$
Q=\sum_i q_i
$$
commutes with the unitary $U_{QCA}$.

Then, we show later that we can ``twist" the QCA by boundary angles $\theta_1,...,\theta_d$, defining a continuous family of QCAs: $U_{QCA}(\theta_1,...,\theta_d)$.  Such a continuous family of QCAs is a continuous map from the $d$-dimensional torus to the unitaries.  The torus of angles $\theta_1,...,\theta_d$ is sometimes called the ``flux torus".
Continuous maps from the torus to the unitaries have been classified.  For $d=1$, they are classified by integers $\ZZ$. For $d=2$, they are classified by $\ZZ \oplus \ZZ$; however, these two integer invariants in $d=2$ are, in a sense, lower dimensional invariants.  They can be computed by considering the dependence upon just one of the two angles.  For $d=3$, we have three lower dimensional integer invariants corresponding to the three different angles, and an additional integer invariant which cannot be obtained from any lower invariant.  In general, recall that the classification of maps from the sphere to the unitaries is given by $\pi_d(U)=Z$ for $d$ odd and $\pi_d(U)=0$ for $d$ even, while the classification of maps from the torus to the unitaries in $d$ dimensions will have some lower dimensional invariants which can be obtained by considering the dependence on only a subset of the angles for all $d\geq 1$ and also will have an integer invariant which cannot be obtained from any lower dimensional invariant for $d$ odd.

We now define the twist.  Consider a QCA $U_{QCA}$ on a torus parameterized by coordinates $x_1,...,x_d$ with $0 \leq x_i < \Ls$.
Unfurl the QCA to a QCA $\alpha$ on the infinite plane, periodic under translation by $\Ls$ in any of the $d$-directions.  Let each site $i$ correspond to a point $x_i$ in this plane, with coordinates $\vec x_i$.  Then, define $\beta(\theta_1,...,\theta_d)$ to be the QCA which conjugates by the unitary $\exp(i \sum_i \frac{q_i \vec \theta \cdot \vec x_i}{\Ls}) $.  Define
\be
\alpha(\theta_1,...,\theta_d)=\beta \circ \alpha \circ \beta^{-1}.
\ee
Because $Q$ commutes with $U_{QCA}$, one can show that $\alpha(\theta_1,...,\theta_d)$ is still periodic.  Hence, one can ``furl" this QCA.  That is, there is a QCA on the torus whose unfurling is $\alpha(\theta_1,...,\theta_d)$.  This defines a unitary $U_{QCA}(\theta_1,...,\theta_d)$ up to a phase ambiguity.
In dimension $d>1$, we fix the phase ambiguity in an arbitrary way, for example by keeping the determinant constant as a function of $\theta_i$.
This give candidate invariants of QCAs on the torus with a continuous symmetry.  We have not constructed QCA with nontrivial values of these invariants; it may be that we need to consider instead LPU with some exponentially decaying control function to find such, which will require some extension of the furling and unfurling.

In $d=1$, the phase of the unitary $U(\theta)$ is essential to defining the invariant.  We could work to unambiguously define the phase but we already have defined an invariant of one-dimensional QCAs with symmetry above so we do not do this.

\section{Discussion}
We have used the torus trick to classify quantum phases.  The general use is to reduce the problem of classifying aperiodic systems to that of classifying periodic systems.
It is not possible to use this trick for all systems.  First, we need to be able to define a pullback.  We can pullback Hamiltonians but we cannot necessarily pullback wavefunctions.  However, even if we pullback a Hamiltonian, if there is intrinsic topological order then it may not be possible to heal the puncture.

Here is one possible further use (we hope there will be more): given gapped two translationally invariant Hamiltonians obeying conditions TQO-1,TQO-2 in Ref.~\onlinecite{bravyihastings}, connected by a gapped path of local Hamiltonians $H_s$, is there a gapped path of translationally invariant locally Hamiltonians connecting them?  We can show that this is true (under one technical assumption) by using quasi-adiabatic continuation to define a path of LGU mapping the ground state of the first Hamiltonian to the second.  Suppose this path of LGU in fact is a path of quantum circuits (this is the technical assumption).  Then, we can apply the torus trick to construct a path of translationally invariant LGUs, $\alpha_s$, connecting the ground state of the first Hamiltonian to the second (the condition TQO-2 is used here in ``healing the puncture").  Then, consider a path $\alpha_s(H_0)$ from $s=0$ to $s=1$ followed by linear interpolation from $\alpha_1(H_0)$ to $H_1$.  This will be explained in more detail elsewhere.

We have given a very brief discussion of the classification of QCA with and without symmetry.  This is a matter for future work.  Also in future work we will consider further the problem of classifying periodic systems without intrinsic topological order.

A very interesting open question is whether there are non-trivial invariants for QCA without symmetry in $d>1$.  This will also be discussed elsewhere.

{\it Acknowledgments---} I thank M. H. Freedman for explaining Kirby's torus trick, and K. Walker for asking about translationally non-invariant paths between Hamiltonians, and  Z. Wang for very useful discussions on cohomology.

\appendix
\section{Healing the Puncture for Free Fermions}
\label{heal}
Here we provide some details on how to heal the puncture for free fermions, giving a simple construction.

We consider a vector space which is a direct sum of two vector spaces of dimensions $N_1,N_2>0$.
We sometimes write vectors $\psi=(\psi_1,\psi_2)$.
The notation here indicates that the first $N_1$ entries of $\psi$ are given by $\psi_1$ which is a vector with $N_1$ components and the next $N_2$ entries by $\psi_2$ which is a vector with $N_2$ components.
The space of dimension $N_2$ corresponds to the sites near the puncture while the space of dimension $N_1$ corresponds to those far from the puncture.

Let $H$ be a matrix
\be
H=\begin{pmatrix}
A & B \\
B^\dagger & 0
\end{pmatrix},
\ee
where
$A$ is an $N_1$-by-$N_1$ Hermitian matrix, $B$ is an $N_1$-by-$N_2$ matrix, and $0$ is the $N_2$-by-$N_2$ zero matrix.

Assume that for any
$\psi=(\psi_1,0)$ that
\be
\label{assume}
|H \psi | \geq c |\psi|
\ee
for some $0<c<1$.

We will show
\begin{theorem}
\label{sup}
For any $N_2$-by-$N_2$ Hermitian matrix $C$, define
\be
J=\begin{pmatrix}
A & B \\
B^\dagger & C
\end{pmatrix}.
\ee
Let $\lambda_{min}(C)$ denote the smallest eigenvalue of $J$ in absolute value.
Then, the supremum of $\lambda_{min}(C)$ over all $C$ is greater than or equal to $c$.
\end{theorem}
The above theorem will not give a norm bound on $C$; we will give a version with a norm bound later after proving this.

Since the supremum might not be achieved for any given $C$, we define a generalization that allows us to work over a compact set.  Let $U$ be a unitary and let $\Lambda$ be a diagonal matrix whose entries are chosen from
the real line union a point at infinity.  Then, we define
$\lambda_{min}(U,\Lambda)$ as follows: given any sequence of $U_i,\Lambda_i$ with $U_i \rightarrow U$ and with $\Lambda_i\rightarrow \Lambda$ with the entries of $\Lambda_i$ chosen real and finite, then define
$\lambda_{min}(U,\Lambda)$ to be the limit of $\lambda_{min}(U_i^\dagger \Lambda_i U_i)$.
This limit exists and is independent of the sequence $U_i,\Lambda_i$ and is a continuous function of $U,\Lambda$ as we will show.

If all entries of $\Lambda$ are finite, these properties are immediate since $\lambda_{min}(C)$ is a continuous function of $C$ and in this case $\lambda_{min}(U,\Lambda)=\lambda_{min}(U^\dagger \Lambda U)$.
To show continuity near infinity, we consider a small neighborhood of some $U_0$ and some $\Lambda_0$ which has some number of finite and some number of infinite entries.
For any $U,\Lambda$ in this neighborhood,
we have a sequence
\be
J_i=\begin{pmatrix}
A & B_{i,\perp} & B_{i,\parallel} \\
B_{i,\perp}^\dagger & C_i^< & 0\\
B_{i,\parallel}^\dagger & 0 & D_i
\end{pmatrix},
\ee
where $C_i^<,D_i$ are diagonal matrices corresponding to finite entries and infinite entries of $\Lambda_0$ respectively (i.e., the limit of each entry of $C_i^<$ is in some small neighborhood of a finite value and the limit of each entry of $D_i$ is in some small neighborhood of infinity, so we will assume that for all sufficiently large $i$, the entries of $D_i$ are all at least equal to some $\lambda_{big}$ in absolute value).  Here the
block matrix
$\begin{pmatrix} B_{i,\perp} & B_{i,\parallel} \end{pmatrix}$ is equal to $B U_i^\dagger$.

In general, if $J_i$ has an eigenvalue $\lambda$ and $\lambda$ is not an eigenvalue of $D_i$, then
the matrix
\begin{eqnarray}
\begin{pmatrix}
A-B_{i,\parallel} (D_i-\lambda)^{-1} B_{i,\parallel}^\dagger \quad& B_{i,\perp} \\
B_{i,\perp}^\dagger & C_i^<
\end{pmatrix}
&=&
\begin{pmatrix}
A-B_{i,\parallel} D_i^{-1} B_{i,\parallel}^\dagger \quad& B_{i,\perp} \\
B_{i,\perp}^\dagger & C^<
\end{pmatrix}+
\begin{pmatrix}
B_{i,\parallel} (D_i^{-1}-(D_i-\lambda)^{-1}) B_{i,\parallel}^\dagger &0\\
0 & 0
\end{pmatrix}
\end{eqnarray}
also has an eigenvalue $\lambda$.  
This method where we compute eigenvalues of a matrix from eigenvalues of a smaller matrix will be used again later; we called it a self-energy technique.
So, if the smallest eigenvalue of $D_i$ is equal to $\lambda_{big}$,
the eigenvalues of $J_i$ in any bounded interval are also eigenvalues of
$$J^{eff}_i \equiv \begin{pmatrix}
A-B_{i,\parallel} D_i^{-1} B_{i,\parallel}^\dagger \quad& B_{i,\perp} \\
B_{i,\perp}^\dagger & C_i^<
\end{pmatrix}$$
up to $\mO(1/\lambda_{big}^2)$ corrections.
Thus, in the given neighborhood, if we define
$$J^{eff}(U,\Lambda) \equiv \begin{pmatrix}
A-B_{\parallel} D^{-1} B_{\parallel}^\dagger \quad& B_{\perp} \\
B_{\perp}^\dagger & C^<
\end{pmatrix},$$
where $B_{\parallel},B_{\perp},D,C^<$ are limits of $B_{i,\parallel},B_{i,\perp},D,C^<_i$, respectively,
then $\lambda_{min}(U,\Lambda)$ is the smallest eigenvalue of $J^{eff}(U,\Lambda)$ in absolute value up to
$\mO(1/\lambda_{big}^2)$ corrections (we have a bound on the absolute value of $\lambda_{min}$ since it is at most $\Vert A \Vert + \Vert B \Vert$).
Hence, for every $U,\Lambda$ in this neighborhood, $\lambda_{min}(U,\lambda)$ is within
$\mO(1/\lambda_{big})$ of the smallest eigenvalue of
$$J^{eff}(U,\Lambda) \equiv \begin{pmatrix}
A& B_{\perp} \\
B_{\perp}^\dagger & C^<
\end{pmatrix},$$
which is obviously continuous in $U,C^<$.

From continuity and compactness of the set of possible $U,\Lambda$, it follows that the supremum is achieved for some
  $U_0,\Lambda_0$.
We will study how $\lambda_{min}$ changes under small changes in $U,\Lambda$. We now change the parametrization of $U,\Lambda$. 
Any change in $U$ that preserves the block structure of $J_i$ (i.e., any change that maps the second block into the second block and the third block into the third block) can be absorbed into a change in $C^<,D^{-1}$ where now we allow them to be arbitrary Hermitian matrices.
 Thus, in the given neighbhorhood we consider matrices
 $$
J_{eff}(U,\Lambda)= \begin{pmatrix}
A-B_{\parallel} D^{-1} B_{\parallel}^\dagger \quad& B_{\perp} \\
B_{\perp}^\dagger & C^<
\end{pmatrix}$$
with $C^<,D^{-1}$ arbitrary Hermitian matrices, with $D^{-1}$ in the neighborhood of the zero matrix (where the supremum is assumed to occur) and $C^<$ in the neighborhood of some matrix $C^<_0$ (where the supremum is assume to occur).
There are also change in $U$ that changes $B_{\parallel},B_{\perp}$ but we will not need to consider these.

Let the smallest eigenvalue of $J_{eff}(U_0,\Lambda_0)$ in absolute value be $\lambda$.  Let $v(1),\ldots,v(k)$ be an orthogonal basis for the eigenectors with eigenvalue $\pm \lambda$.  Let eigenvector $v(a)$ have eigenvalue $\sigma(a) \lambda$ where $\sigma(a)=\pm 1$.
Let $\Pi_2$ project onto the second block of matrix $J^{eff}$.
If the vectors $\Pi_2 v(1),\ldots,\Pi_2 v(k)$ are linearly independent, then we can choose vectors $w(1),\ldots,w(k)$ so that $\langle w(a) | \Pi_2 v(b) \rangle =\delta_{a,b}$.  Then, by setting $C^<=C^<_0+\epsilon \sum_a \sigma(a) |w(a) \rangle \langle w(a) |$ for small $\epsilon$ this increases the minimum eigenvalue of $J_{eff}$ in absolute value. 
Hence, the vectors $\Pi_2 v(a)$ must be linearly dependent.
This means that some vector $v$ which is a linear combination of $v(a)$ has $\Pi_2 v=0$.  
Consider the space of such vectors $v$.  Let $\Pi_1$ project onto the first block of matrix $J^{eff}$.  We claim that at least one such vector
must also having $\Pi_1 v$ vanishing in the range of $B_{\parallel}$, otherwise by choosing a small $D^{-1}$ we could increase the eigenvalue in absolute value.  Hence, $B_{\parallel}^\dagger \Pi_1 v=0$.  So, $$|J_{eff}(U_0,\Lambda_0) v|=\sqrt{|A \Pi_1 v|^2+|B_{\perp}^\dagger \Pi_1 v|^2}=\sqrt{|A \Pi_1 v|^2+|B^\dagger \Pi_1 v|^2} \geq c|v|$$ where the inequality is by Eq.~(\ref{assume}).
This completes the proof of the theorem.

To clarify this construction, consider two examples.  First, consider the case that
$N_1=N_2=1$ and
$$H=\begin{pmatrix} 0 & c \\ c & 0\end{pmatrix}.$$
If $C$ is a scalar $z$ then we have
$$J=\begin{pmatrix} 0 & c \\ c & z\end{pmatrix},$$
with the maximum  occuring for $z=0$, with eigenvalues $\pm 1$ and the vector $(1,0)$ being a linear combination of the corresponding eigenvectors.

Next consider the case that
$N_1=2,N_2=1$ and
$$J=\begin{pmatrix} 0 & c & 0 \\ c & 0 & 1 \\ 0 & 1 & z\end{pmatrix},$$
where $C$ is a scalar $z$.  Then, maximize by taking $z\rightarrow \infty$.

Now we show how to bound $\Vert C \Vert$, at the cost of also ``removing and adding sites", defined below.

First consider the case with removing sites, not adding them.
We define
``removing sites"
as follows: given an $N_2$-by-$N_2$ projector $Q$,
consider the matrix $H$ projected into the range of $I \oplus (1-Q)$, where $I$ is the $N_1$-by-$N_1$ identity matrix.
Choosing a basis for the range of $1-Q$, we can write this matrix in the form
$$
H'=\begin{pmatrix}
A & B_{\perp}\\
B_{\perp}^\dagger & 0
\end{pmatrix}.
$$
Remark: the effect of removing sites is very similar to taking eigenvalues to infinity in $\Lambda$.

We now define ``adding sites" by considering a matrix
$$
H''=\begin{pmatrix}
A & B_{\perp} & 0\\
B_{\perp}^\dagger & 0 & 0 \\
0 & 0 & 0
\end{pmatrix},
$$
where the matrix is still square but we have padded with an arbitrary number of zero rows and columns.
Now we show that
\begin{theorem}
Assume $\Vert B \Vert \leq 1$.
For any Hermitian matrix $C$ of the appropriate dimensions given by
$$C=\begin{pmatrix}
C_{11} & C_{12} \\ C_{21} & C_{22},
\end{pmatrix}$$
let
\be
\label{nowJ}
J=\begin{pmatrix} A & B_{\perp} & 0 \\
B_{\perp}^\dagger & C_{11} & C_{12}\\
0 & C_{21} & C_{22}
\end{pmatrix}.
\ee
Note that if we do not add any sites, but we do remove sites, then this is the same construction as taking some infinite entries in $\Lambda$ above.

Let $\lambda_{min}(C)$ denote the smallest eigenvalue of $J$ in absolute value.
Then, for any $d>0$, there is some choice of $Q,C$ such that
$\lambda_{min}(C)\geq {\rm const.} \times c+\mO(c^2)$, with the constants depending on $d$ but not on $c$.
\end{theorem}

We show this theorem as follows.
First, using theorem \ref{sup}, find a choice of $C_0$ that gives $\lambda_{min}(C_0)\geq 0.9 c$.  (The constant $0.9$ is of course arbitrary, any number in $(0,1)$ would suffice).
Let $J_0$ be the matrix
$\begin{pmatrix} A & B \\ B^\dagger & C \end{pmatrix}$.

``Remove" all eigenvalues larger than $10 c^{-1}$ in absolute value.  That is, choose $Q$ to project onto the eigenspace of $C_0$ with eigenvalue larger than $10 c^{-1}$ in absolute value, and define a matrix $C_0^<$ on the range of $1-Q$ which is equal to $C_0$ projected onto that range.

Now we show that
$$J^<\equiv \begin{pmatrix} A & B_{\perp} \\ B_{\perp}^\dagger & C_0^< \end{pmatrix}$$
still has smallest eigenvalue at least equal to
$0.8 c -\mO(c^3)$
in absolute value.
Then, for any $\omega$, if $\omega$ is an eigenvalue of $J_0$ and if $\omega$ is not an eigenvalue of $C_{large}$, then
$\omega$ is an eigenvalue of
$$
\begin{pmatrix} A-B_{\parallel} (C^{large}-\omega)^{-1} B_{\parallel}^\dagger \quad & B_{\perp} \\ B_{\perp}^\dagger & C_0^< \end{pmatrix},$$
using the self-energy technique.
Since $\Vert B_{\parallel} \Vert \leq 1$, and $\Vert (C^{large}-\omega)^{-1} \Vert \leq 0.1 c+\mO(c^3)$ for $|\omega|<c$,
this means that
$\omega$ is within $0.1 c+\mO(c^3)$ of an eigenvalue of $J_0$ and since the smallest eigenvalue of $J_0$ is at least $0.9 c$ in absolute value, the smallest eigenvalue of $J^<$ is at least $0.8 c -\mO(c^3)$
in absolute value.

This gives a norm bound on $C_0^<$ but the norm bound is $10 c^{-1}$.
Now ``add sites" as follows.
For every eigenvalue $\lambda$ of $C_0^<$ larger than $d$ in absolute value, add a single site (i.e., add a row and column to the matrix),
and replace
$$
\lambda \rightarrow \begin{pmatrix} 0 & d/2 \\ d/2 & -\frac{d^2}{4\lambda} \end{pmatrix}.$$
This notation means that we add a single row and column to the matrix, we replace the diagonal entry of $C_0^>$ which equals $\lambda$ with a zero entry, and we add a new diagonal entry in the added row and column equal to $-(4 \lambda)^{-1}$, as well as adding two off-diagonal entries equal to $d/2$.
For example, applied to the matrix
$$\begin{pmatrix} \lambda_1 & 0 \\ 0 & \lambda_2 \end{pmatrix},$$ if $|\lambda_1|,|\lambda_2|>d$, then
we replace with
$$\begin{pmatrix} 0 & 0 & d/2 & 0 \\ 0 & 0 & 0 & d/2 \\ d/2 & 0 & -\frac{d^2}{4\lambda_1} & 0 \\ 0 & d/2 & 0 & -\frac{d^2}{4\lambda_2}\end{pmatrix}.$$
Note that $|4 \lambda|^{-1} \geq (2/5) c$.
This construction is inspired by the same kind of self-energy technique as we have $-(d/2)^2/(d^2/4\lambda-\omega) \approx \lambda$ for small $\omega$.

Let $C$ denote the matrix resulting from this construction of removing and adding sites.
We claim that this has the desired properties.  By construction, the matrix $C$ has norm bounded by $d$.
To show the bound on $\lambda_{min}$,
note that if $\omega$ is an eigenvalue of $J$, with $\omega$ not equal to $-d^2/4\lambda$ for any of the added sites, then
$\omega$ is an eigenvalue of a matrix $J^<_{eff}$
that we define as follows, again using the self-energy technique.  First, replace each diagonal
entry $\lambda$ of $C_0^<$ which is larger than $d$ with
$(d/2)^4 (d^2/(4\lambda)-\omega)^{-1}=\lambda/(1-4\omega\lambda/d^2)$ and call the resulting matrix
$C_{eff}^<$.
Then, write
$$J^<_{eff}=
 \begin{pmatrix} A & B_{\perp} \\ B_{\perp}^\dagger & C_{eff}^> \end{pmatrix}.$$

Now we repeat this self-energy technique.  For any eigenvalue $\omega$ of $J^<_{eff}$ which is not an eigenvalue of $C^<_{eff}$, we have that $\omega$ is
an
eigenvalue of
$A-B_{\perp} (C^<_{eff}-\omega)^{-1} B_{\perp}^\dagger$
while for any
 any eigenvalue $\omega$ of $J^<$ which is not an eigenvalue of $C^<$, we have that $\omega$ is
 an
eigenvalue of
$A-B_{\perp} (C_0^<-\omega)^{-1} B_{\perp}$.
One may bound
$\Vert (C^<_{eff}-\omega)^{-1} - (C_0^<-\omega)^{-1} \Vert$ by
the maximum of
$$|(\lambda/(1-4\omega\lambda/d^2)-\omega)^{-1}-(\lambda-\omega)^{-1}|$$
over $\lambda$ in the range $[d,10c^{-1}]$.  For $\omega$ suffciently small compared to $c$, this maximum is bounded by a constant times $c$, with the constant tending to zero as $\omega/c\rightarrow 0$.  This gives
a bound on
$\Vert B_{\perp} (C^<_{eff}-\omega)^{-1} B_{\perp}^\dagger- B_{\perp} (C_0^<-\omega)^{-1} B_{\perp}^\dagger\Vert$ for small $\omega$.

At this point we are almost done; if the quantity $A-B_{\perp} (C_0^<-\omega)^{-1} B_{\perp}^\dagger$ were independent of $\omega$, then it would follow straightforwardly: an eigenvalue
of $A-B_{\perp} (C^<_{eff}-\omega)^{-1} B_{\perp}^\dagger$ which is sufficiently small compared to $c$
would imply a small eigenvalue of $A-B_{\perp} (C_0^<-\omega)^{-1} B_{\perp}^\dagger$ (i.e., an eigenvalue smaller than our proven lower bound on the eigenvalue of that matrix) because the difference in matrices is small.

However, $A-B_{\perp} (C_0^<-\omega)^{-1} B_{\perp}^\dagger$ is not independent of $\omega$, but this does not lead to any difficulty:
if $A-B_{\perp} (C^<_{eff}-\omega)^{-1} B_{\perp}^\dagger$ has an eigenvalue $\omega$ for sufficiently small $\omega$
with eigenvector $\psi_1$, then
the vector
$\psi =(\psi_1,(C_0^<-\omega)^{-1} B_{\perp} \psi_1)$ is such that $|J^< \psi|/|\psi|$ is small compared to $c$, giving a contradiction. 
This completes the proof.

\end{document}